\documentclass[prd,preprint,eqsecnum,amsmath,amssymb,nofootinbib,superscriptaddress]{revtex4}
\usepackage{graphicx, tikz}
\usepackage{amssymb,amsmath, mathrsfs}
\usepackage{amssymb, graphics, setspace}
\usepackage{hyperref}
\usepackage[all]{xy}
\newcommand{\be}{\begin{equation}}
\newcommand{\ee}{\end{equation}}
\newcommand{\bea}{\begin{eqnarray}}
\newcommand{\eea}{\end{eqnarray}}
\newcommand{\bes}{\begin{subequations}}
\newcommand{\ees}{\end{subequations}}

\newcommand{\w}{\omega}
\begin{document}
\title{Correlations between a Hawking particle and its partner in a 1+1D Bose-Einstein condensate analog black hole }
\author{Richard~A.~Dudley}
\email{dudlra13@wfu.edu}
\affiliation{Department of Physics, Wake Forest University, Winston-Salem, North Carolina 27109, USA}

\author{Alessandro~Fabbri}
\email{afabbri@ific.uv.es}
\affiliation{Departamento de F\'isica Te\'orica and IFIC, Universidad de Valencia-CSIC, C. Dr. Moliner 50, 46100 Burjassot, Spain and\\
Universit\'e Paris-Saclay, CNRS/IN2P3,
IJC Lab, 91405 Orsay Cedex, France
}

\author{Paul~R.~Anderson}
\email{anderson@wfu.edu}
\affiliation{Department of  Physics, Wake Forest University, Winston-Salem, North Carolina 27109, USA}

\author{Roberto~Balbinot}
\email{balbinot@bo.infn.it}
\affiliation{Dipartimento di Fisica dell'Universit\`a di Bologna and INFN sezione di Bologna, Via Irnerio 46, 40126 Bologna, Italy\\
}
\clearpage \newpage

\begin{abstract}
The Fourier transform of the density-density correlation function in a Bose-Einstein condensate (BEC) analog black hole is a useful tool to investigate correlations between the Hawking particles and their partners.
It can be expressed in terms of $\langle ^{\text{out}}\hat{a}_{\text{up}}^{\text{ext}}\ ^{\text{out}}\hat{a}_{\text{up}}^{\text{int}} \rangle$, where $^{\text{out}}\hat{a}_{\text{up}}^{\text{ext}}$ is the annihilation operator for the Hawking particle and $^{\text{out}}\hat{a}_{\text{up}}^{\text{int}} $ is the corresponding one for the partner. This basic quantity is calculated for three different models for the BEC flow. It is shown that in each model the inclusion of the effective potential in the mode equations makes a significant difference.
Furthermore, particle production induced by this effective potential in the interior of the black hole is studied for each model and shown to be nonthermal. An interesting peak that is related to the particle production and is present in some models is discussed.
\end{abstract}
\maketitle
\section{Introduction}
Hawking's 1974 prediction\cite{Haw1975199220} that black holes evaporate has not been directly verified, largely because a  black hole of mass $M$ would emit radiation  at a temperature $T_H\sim  \frac{M_\odot}{M}\times 10^{-7}K$. Some hope remains for a detection from  black holes nearing the end of the evaporation process, but ``primordial'' black holes, which formed in the early universe, have not been detected and there is no evidence for radiation from them\cite{PanLoe2014124}.

It was shown in \cite{Unruh1351211981} that a fluid flowing from a subsonic into a supersonic region, and thus having an acoustic horizon,  should also emit a thermal spectrum of phonons via the Hawking effect and therefore serve as an analog black hole. Even in analog systems the temperature  of the emission is usually very low. Bose-Einstein condensates (BECs) have been particularly useful as analog black holes because they are suited for testing low energy phenomena as they can be cooled to ${10^{-7}\text{K}}$\cite{Leanhardt1513}. These systems can be effectively trapped in a one-dimensional (1D) flow, creating an analog spacetime with 1+1 dimensions.  Direct detection of the produced phonons is still problematic; therefore, other signatures of the Hawking process are the focus of current quantum field theory in curved space predictions and analog black hole experiments.

The most notable prediction associated with the Hawking effect in analog systems to date has been a peak in the correlation function for the density in a 1+1D BEC analog black hole. This prediction was originally made using quantum field theory in curved space for a simple model with  a constant flow velocity and a varying sound speed\cite{PhysRevA.78.021603}. It was subsequently verified by a quantum mechanics calculation\cite{CarFag2008103001, PhysRevA.80.043603} and a more sophisticated quantum field theory in curved space calculation\cite{AndBal1240182013}.

Experiments using a 1+1D BEC analog black hole  in 2016\cite{Ste2016959} and 2019\cite{Nature.569.s41586} found very good qualitative agreement with the prediction of the peak in the density-density correlation function. These experiments have position-dependent sound speeds and flow velocities in an effectively one-dimensional system.  The density for all points in each experimental run is imaged at one lab time. The experiment is repeated several thousands of times to build an ensemble average for the density-density correlation function. The peak predicted by the constant flow velocity model is clearly evident in the experimental results.

An attempt was made to model the 2016 experiment in \cite{2016PhRvD94h4027M}. The model uses a step function potential to obtain an analytic solution to the Gross-Pitaevskii equation which governs the background condensate. Several quantities were calculated including the density-density correlation function.  An approximation was used in the calculation for the density-density correlation function that involved setting an effective  potential that appears in the phonon mode equation to zero. When the cross section of the resulting density-density correlation function was compared to the experimental result, there was nearly a factor of  2 difference in the  full width half maximum of the peak and $\sim50\%$ difference for the height of the peak.

In order to determine the temperature of the analog black hole the experimenters decomposed the peak found in the density-density correlation function via a Fourier transform to show the correlation spectrum for the Hawking particle and its partner\cite{Nature.569.s41586}. In \cite{PhysRevD.92.024043} a theoretical quantity, which we call the Hawking-partner  (HP) correlator, was shown to be related to this Fourier transform. In \cite{Nature.569.s41586} the spectrum of the HP correlator was calculated using an approximation in which the effective potential in the phonon mode equation was ignored. In this case, the HP correlator only depends on the frequency of the modes and the surface gravity, and hence the temperature of the analog black hole. A comparison was made with the experimental result.  A disagreement of  $ 1\% $ was found for the temperature of the analog black hole. The authors estimated an experimental error in this quantity of $5\%$.  The effects of nonlinear dispersion on this correlator were investigated in \cite{PhysRevLett.124.060401}, but this calculation did not include an effective potential in the phonon mode equation.

In this paper, we will work with three different models for a 1+1D BEC analog black hole. We calculate the resulting HP correlator and two quantities related to the population of phonons traveling upstream and downstream in the frame of the fluid in the interior of the acoustic black hole, and we find  that there is a significant contribution from the effective potential for each model to all of these quantities.

The first model, previously discussed in \cite{PhysRevA.80.043603,PhysRevD.100.105021}, has an effective potential consisting of  a delta function in the interior and  a delta function in the exterior of the BEC analog black hole. This model is simple enough so that analytic results were obtained.   We compare to the case with no potential and thus no scattering or particle production and find significant differences.

We then review the profile used in \cite{CarFag2008103001,AndBal1240182013}, which has a varying speed of sound and a constant flow velocity. The effective potential is included in the mode equation and we find that the HP correlator, again, is significantly altered by its inclusion.

The third model we look at is often called the waterfall model~\cite{ PhysRevA.68.063608,PhysRevA.80.043603,PhysRevA.85.013621,2016PhRvD94h4027M}. It has been used to model the 2016 experiment in \cite{2016PhRvD94h4027M}.  Here the term waterfall refers to an analytic solution to the Gross-Pitaevskii equation for a BEC analog black hole in which the condensate is flowing over a step function potential.  In this model, the sound speed, flow velocity and background density all vary along the flow direction.
 In this case, we also find that the HP correlator is significantly altered due to the scattering and particle production caused by the inclusion of the effective potential.

We then discuss a new peak found in \cite{PhysRevD.100.105021} related to the population of phonons propagating upstream and downstream in the frame of the fluid inside the horizon, for each of these models. We will refer to these as the interior upstream phonon number(UPN) and the interior downstream phonon number (DPN).  This peak was found to occur when the magnitude of the effective potential is larger in the interior than in the exterior. The peak  was noted previously for the two-delta function potential in \cite{PhysRevD.100.105021} when the interior potential is chosen to be larger than the exterior.  The second profile, which has a constant flow velocity and an effective potential whose magnitude is similar in the interior and exterior regions, exhibits no such peak.  The last model we investigate  is the waterfall model  which displays a relatively large peak in quantities related to the population of phonons in the interior.

In Sec.~\ref{Sec:HpCorrelatorBackground}, we discuss the theoretical background for a 1+1D BEC analog black hole.  We then derive the HP correlator based on the creation and annihilation operators for a Hawking phonon and its partner for the two-delta function potential model. We also compute the HP correlator when the effective potential is ignored.  In Sec. \ref{Sec:HPCorrelatorWithPotential}, first the constant flow velocity model is reviewed and our results for the HP correlator are given.   Then the waterfall model is reviewed and our results for the HP correlator for it are shown. In Sec.~\ref{Sec:InteriorParticleProduction3Models},   we discuss the overall effect of the potential in each case on the appearance of the peak which is related to particle production in the interior. In Sec.~\ref{Sec:HpCorrelatorConclusions}, we discuss our results.

\clearpage
\section{Background}
\label{Sec:HpCorrelatorBackground}
The field equation for the phonon operator $\hat\theta_1$,
 if the flow velocity, $\vec{v}$, sound speed, $c$, and density, $n$, change
on scales larger than the healing length\footnote{ The healing length sets the scale of dispersion.}  $\xi=\frac{\hbar}{mc}$, with $m$ the mass of an atom,
is (see e.g.~\cite{Visserlrr-2005-12})
\be
\left[-\left(\partial_T+\vec{\nabla}\cdot\vec{v}\right)\frac{n}{mc^2}\left(\partial_T+\vec{\nabla}\cdot\vec{v}\right)+\vec{\nabla}\cdot \frac{n}{m}\vec{\nabla}\right]\hat{\theta}_1=0.\label{Eq:2013PRD}
\ee
  The coordinates $T$ and $\vec{x}$ relate to the lab frame. Equation~\eqref{Eq:2013PRD} is equivalent to the Klein-Gordon equation for a massless scalar field in a curved spacetime with line element  of the form
\be
ds^2=\frac{n}{mc}\left[-(c^2)dT^2+(d\vec{x}-\vec{v}dT)\cdot(d\vec{x}-\vec{v}dT)\right]
\ee
We consider flows that are stationary and effectively one-dimensional and we define a 1+1D field operator, $\hat{\theta}^{(2)}$, such that
\be
 \hat{\theta}_1 = \sqrt{\frac{mc(x)}{n(x)\hbar l_\perp^2}}\hat{\theta}^{(2)} \;,
 \ee
 where $l_\perp$ is a length, defined by the transverse confinement of the BEC.  For analog black holes the condensate is flowing from a subsonic region
 $c>|\vec{v}|$ ($x>0$, region r)  into a supersonic region with $c<|\vec{v}|$ ($x<0$, region l). For the models considered in this paper the flows are directed from $x=\infty$ to $x=-\infty$, so $\vec{v}=-v_0(x)\hat{x}$, with $v_0>0$. The condensates we consider also have the property that they either have or approach a constant flow velocity, sound speed, and density as $x\to \pm \infty$. In the analog spacetime this translates to a region that is asymptotically flat.
Using the variable transformations
\be
t=T-\int^xdy \frac{v_0(y)}{c(y)^2-v_0(y)^2}+a \quad \text{and} \quad x^*=\int^{{x}}dy\frac{c(y)}{c(y)^2-v_0(y)^2}+b
\ee
with $a$ and $b$ arbitrary constants,\footnote{It is useful to fix the constants $a$ and $b$ so that $\text{v}=t+x^*$ is continuous across the horizon.} the  equation for $\hat{\theta}^{(2)}$ is
\be
\left(-\partial_t^2 +\partial_{x^*}^2+V_{\text{eff}} \right)\hat{\theta}^{(2)}=0, \label{Eq:KGEMMSFNCVofX}
\ee
with the effective potential
\bea
V_{\text{eff}}=\frac{v_0^4 }{2 c^3}\frac{d^2c}{dx^2}-\frac{v_0^2 }{c}\frac{d^2c}{dx^2}+\frac{1}{2} c \frac{d^2c}{dx^2}+\frac{v_0^4 }{c^3 n}\frac{dc}{dx} \frac{dn}{dx}-\frac{v_0^2 }{c n}\frac{dc}{dx}  \frac{dn}{dx}+\frac{v_0^3 }{c^3}\frac{dc}{dx} \frac{dv_0}{dx}-\frac{v_0 }{c}\frac{dc}{dx} \frac{dv_0}{dx}\nonumber \\ -\frac{5 v_0^4 }{4 c^4}\left(\frac{dc}{dx} \right)^2+\frac{3 v_0^2 }{2 c^2}\left(\frac{dc}{dx} \right)^2-\frac{1}{4} \left(\frac{dc}{dx} \right)^2-\frac{v_0^4 }{2 c^2 n}\frac{d^2n}{dx^2}-\frac{c^2 }{2 n}\frac{d^2n}{dx^2}-\frac{v_0^3 }{c^2 n}\frac{dn}{dx} \frac{dv_0}{dx}\nonumber \\+\frac{v_0^4 }{4 c^2 n^2}\left(\frac{dn}{dx}\right)^2 +\frac{c^2 }{4 n^2}\left(\frac{dn}{dx}\right)^2+\frac{v_0^2 }{n}\frac{d^2n}{dx^2}+\frac{v_0 }{n}\frac{dn}{dx} \frac{dv_0}{dx}-\frac{v_0^2 }{2 n^2}\left(\frac{dn}{dx}\right)^2.\ \ \label{effpot}
\eea
Note that $v_0$ and $n$ are related by the continuity equation $nv_0= $ const.
 The asymptotically constant flows we are considering ensure that the effective potential vanishes in the limit $x\to \pm\infty$. It also vanishes on the horizon, $x=0$.
 The wave equation~\eqref{Eq:KGEMMSFNCVofX} can be written in the form $ (\Box^{(2)} + V) \hat{\theta}^{(2)} = 0$ where $\Box^{(2)}$ is the d'Alambertian for the two-dimensional metric
\be
ds^2=\frac{ \left[c(x)^2-v_0(x)^2\right]}{c(x)}\left(-dt^2+dx^{*2}\right), \label{LineElementGeneral}
\ee
and $V = c\, (c^2-v_0^2)^{-1} V_{\rm eff}$.
It is useful to define the ingoing and outgoing null coordinates $\text{v}=t+x^*$  and $u=t-x^*$, and the Kruskal null coordinates
\begin{align}
U=\mp\frac{e^{-\kappa u}}{\kappa}\quad \text{and} \quad V=\frac{e^{\kappa \text{v}}}{\kappa}.
\end{align}
 Here the $-$ and $+$ refer to the exterior and interior regions, respectively, of the analog  spacetime and the surface gravity, $\kappa$, is defined as
\be
 \kappa = \left. \left(\frac{d c}{dx} - \frac{dv_0}{dx}\right)\right|_{\rm hor}  \;.
\ee

\begin{figure}[h!]	
		\includegraphics[trim=1cm 11cm 0cm 0cm, clip=true, totalheight=0.6\textheight]{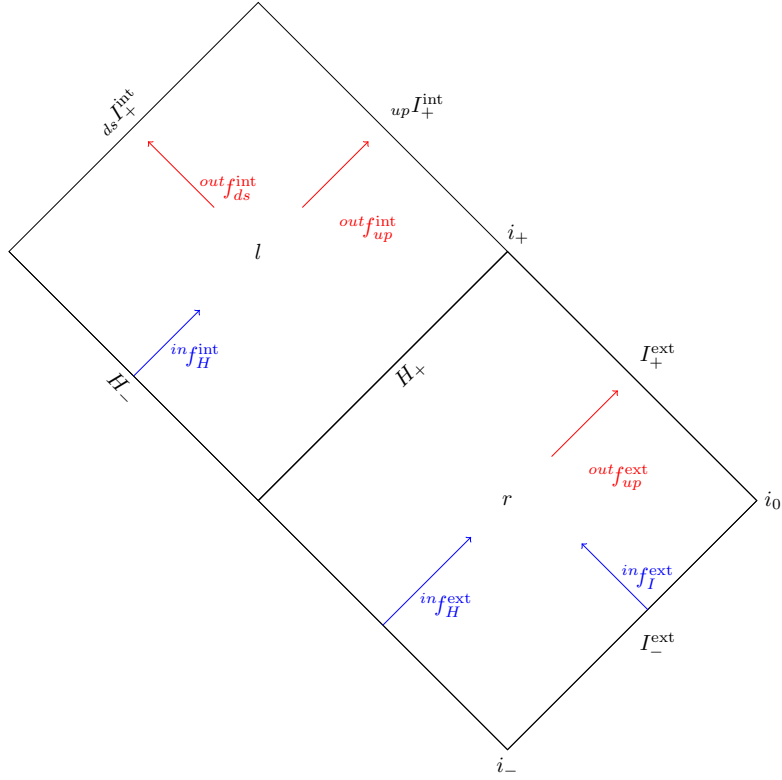}
\caption{\label{Fig:Penrose1} Penrose diagram for an analog spacetime corresponding to a BEC flowing from right to left. The {\it in} mode basis is schematically illustrated in blue  in the ${l}$ and ${r}$ regions of an analog black hole. The {\it out} mode basis is schematically illustrated in red.  }
\end{figure}

In order to proceed, we need to define two quantum states for the field.  These can be described by complete sets of modes that are positive frequency  on certain surfaces.  We start with the Boulware state which is defined by solutions to the mode equation \eqref{Eq:KGEMMSFNCVofX} that are positive frequency with respect to $t$ on $I_-$ and the past horizon $H_-$ in the region outside the future horizon. Inside the future horizon they are positive frequency with respect to the time coordinate $x^*$ on the past horizon.  The Penrose diagram in Fig.~\ref{Fig:Penrose1} helps illustrate the behaviors of these modes in the analog spacetime.
On the past horizon, they take the form
\be
  ^{\text{in}}\!f_H^{\text{ext}}=\frac{e^{-i\omega \text{u}}}{\sqrt{4\pi\omega}}\quad \text{and} \quad   ^{\text{in}}\!f_H^{\text{int}}=\frac{e^{i\omega \text{u}}}{\sqrt{4\pi\omega}}.
\ee
In what follows, we use the superscript ``ext'' to denote modes that are positive frequency on a surface in the exterior region and ``int'' to denote modes that are positive frequency on a surface in the interior region. The subscript $H$ or $I$ denotes whether  that surface is a horizon or null infinity.  The superscript ``in'' denotes the {\it{in}} modes and the superscript ``out'' denotes the {\it{out}} modes.  The modes on $I_-$ have the form
   \be
  ^{\text{in}}\!f_I^{\text{ext}}=\frac{e^{-i\omega \text{v}}}{\sqrt{4\pi\omega}}. \label{Eq:_MinusModes}
  \ee
  Since these modes form a complete orthonormal set, the field can be expanded in terms of them as
\be
\hat{\phi}^{(2)}=\int_0^\infty d\omega \left[^{\text{in}}\hat{a}_H^{\text{ext}} (^{\text{in}}\!f_H^{\text{ext}})+ ^{\text{in}}\hat{a}_H^{\text{int}} (^{\text{in}}\!f_H^{\text{int} })+^{\text{in}}\hat{a}_I^{\text{ext}} (^{\text{in}}\!f_I^{\text{ext}})+\text{H.c.}\right].
\ee
Here  $^{\text{in}}\hat{a}_H^{\text{ext}}$, $^{\text{in}}\hat{a}_H^{\text{int}}$, and $^{\text{in}}\hat{a}_I^{\text{ext}}$ are  annihilation operators and the Boulware vacuum is the state annihilated by these operators.

The Boulware state does not correctly describe the state of the quantum field when the black hole is created dynamically.  In this case, at late times,
the state of the quantum field is well approximated by
the Unruh state~\cite{PhysRevD.14.870}.
The Unruh  state consists of modes that are positive frequency with respect to the Kruskal time coordinate on the past horizon so that
\be
f_H^K=\frac{e^{-i\omega_K U}}{\sqrt{4\pi\omega_K}}
\ee
and modes that are positive frequency with respect to $t$ on $I_-$ shown in Eq. \eqref{Eq:_MinusModes}.  These two sets of modes form a complete orthonormal set and the field can then be expanded in terms of them as
  \be
\hat{\phi}^{(2)}=\int_0^\infty d\omega_K \left[(\hat{a}_{\omega_K}\ f_H^K +\hat{a}_{\omega_K}^\dagger\ f_H^{K*}\right) +\int_0^\infty d\omega \left[^{\text{in}}\hat{a}_I^{\text{ext}}\ (^{\text{in}}\!f_I^{\text{ext}})+\ ^{\text{in}}\hat{a}_I^{\text{ext} \dagger}\ (^{\text{in}}\!f_I^{\text{ext} *})\right]. \label{decompunruh}
\ee
The Unruh $|U\rangle$ state is state annihilated by all the annihilation operators entering the decomposition given in~(\ref{decompunruh}).
Here $\hat{a}_{\omega_K}$ is an annihilation operator for the $f_H^K$  modes.    The mode equation in  Kruskal coordinates is not separable; thus, it is preferable to work with the modes that specify the Boulware state.
The relation between the two sets of annihilation operators is given by the following Bogoliubov transformations
\bea
^{\text{in}}\hat{a}_H^{\text{ext}}=\int_0^\infty d\omega_K \left[\alpha^{\text{ext}}_{\omega_K \omega}\hat{a}_{\omega_K} +\beta^{\text{ext} *}_{\omega_K \omega}\hat{a}_{\omega_K}^\dagger\right]\ , \nonumber \\
^{\text{in}}\hat{a}_H^{\text{int}}=\int_0^\infty d\omega_K \left[\alpha^{\text{int}}_{\omega_K \omega}\hat{a}_{\omega_K} +\beta^{\text{int} *}_{\omega_K \omega}\hat{a}_{\omega_K}^\dagger\right]\ , \label{Eq:IntoKruskalOperators}
\eea

For a late time observer, what we would think of as the natural {\it{out}} vacuum state consists of a complete set of modes that are positive frequency with respect to $t$ or $x^*$, on $I_+$, where $I_+$ refers to the entire surface of future null infinity.  In the exterior region on $I_+^{\text{ext}}$, the upstream modes take the form
\be
 ^{\text{out}}\!f_{\text{up}}^{\text{ext}}=\frac{e^{-i\omega \text{u}}}{\sqrt{4\pi\omega}} \;.
\ee
In the interior region, the upstream modes on the surface $_{\text{up}}\!I_+^{\text{int}}$ are
\be
 ^{\text{out}}\!f_{\text{up}}^{\text{int}}=\frac{e^{i\omega \text{u}}}{\sqrt{4\pi\omega}} \;,
\ee
 and the interior downstream modes on $_{\text{ds}\,}\!I_+^{\text{int}}$
are
 \be
 ^{\text{out}}\!f_{\text{up}}^{\text{int}}=\frac{e^{-i\omega \text{v}}}{\sqrt{4\pi\omega}}.
\ee
 The three surfaces that comprise $I_+$   and the {\it{out}} state are illustrated in Fig. \ref{Fig:Penrose1}. The field can be expanded in terms of these modes as well
\be
\hat{\phi}^{(2)}=\int_0^\infty d\omega \left[ ^{\text{out}}\hat{a}_{\text{up}}^{\text{ext}}\ (^{\text{out}}\!f_{\text{up}}^{\text{ext}}) +\ ^{\text{out}}\hat{a}_{\text{up}}^{\text{int}}\ (^{\text{out}}\!f_{\text{up}}^{\text{int}})+\ ^{\text{out}}\hat{a}_{\text{ds}}^{\text{int}}\ (^{\text{out}}\!f_{\text{ds}}^{\text{int}})+ \text{H.c.}\right]
\ee
where the $^{\text{out}}\hat{a}$'s are the associated annihilation operators.

   In general one can use scattering theory to relate the modes in the  {\it{in}} states to those in the {\it{out}} states.  An exterior {\it{in}} mode  initially propagates downstream away from past null infinity  and is partially reflected upstream toward $I^{\text{ext}}_+$ with a reflection coefficient of  $R_I^{\text{ext}}$. The transmitted portion continues to travel downstream into the interior of the analog black hole where it undergoes particle production.\footnote{The scattering in the interior region is anomalous and this leads to particle production; see~\cite{PhysRevD.100.105021}.} After the particle production occurs, the part of the mode that travels upstream toward $_{\text{up}}I_+^{\text{int}}$  has a total scattering coefficient of $R_I^{\text{int}}$, while the portion of the mode that continues to travel toward $_{\text{ds}}I_+^{\text{int}}$ has a total scattering coefficient $T_I^{\text{int}}$. The other modes have similar behaviors and one can write the {\it{in}} modes
 on $I_+$ in terms of the {\it{out}} modes as follows:
\bes \bea
&^{\text{in}}\!f_I^{\text{ext}}=R_I^{\text{ext}} \  ^{\text{out}}\!f_{\text{up}}^{\text{ext}}+T_I^{\text{int}} \ ^{\text{out}}\!f_{\text{ds
}}^{\text{int}} + R_I^{\text{int}} \  ^{\text{out}}\!f_{\text{up}}^{\text{int} *},\label{Eq:IntoOut1}\\
&^{\text{in}}\!f_H^{\text{ext}}=T_H^{\text{ext}} \ ^{\text{out}}\!f_{\text{up}}^{\text{ext}}+T_H^{\text{int}} \ ^{\text{out}}\!f_{\text{ds}}^{\text{int}} + R_H^{\text{int}} \  ^{\text{out}}\!f_{\text{up}}^{\text{int} *} ,\label{Eq:IntoOut2}\\
&^{\text{in}}\!f_H^{\text{int}}=\quad \quad \quad \quad \quad \ \ \tilde{R}_H^{\text{int}} \ ^{\text{out}}\!f_{\text{ds}}^{\text{int} *} + \tilde{T}_H^{\text{int}} \  ^{\text{out}}\!f_{\text{up}}^{\text{int}}.\label{Eq:IntoOut3}
\eea \ees
 Note that the tilde denotes a coefficient which does not involve any scattering in the exterior region. One can now formulate the scattering matrix and using scattering theory we can then calculate the expressions for the annihilation operators for the {\it out} state in terms of those for the Unruh state,
\begin{subequations}
\begin{align}
^{\text{out}}\!\hat{a}_{\text{up}}^{\text{ext}}=(\hat{\phi}^{(2)},\  ^{\text{out}}\!f_{\text{up}}^{\text{ext}})&=\int_0^\infty d\omega_K\left[\hat{a}_{\omega_K}(\alpha_{\omega_K \omega}^{\text{ext}})T_H^{\text{ext}}+\hat{a}_{\omega_K}^\dagger (\beta_{\omega_K \omega}^{\text{ext} *})T_H^{\text{ext}}\right] \nonumber \\ &\quad \quad+^{\text{in}}\!\hat{a}_I^{\text{ext}}R_I^{\text{ext}} \;,\label{Eq:AnihhilationOperator1} \quad \\
^{\text{out}}\!\hat{a}_{\text{up}}^{\text{int}}=(\hat{\phi}^{(2)},\  ^{\text{out}}\!f_{\text{up}}^{\text{int}})=&\int_0^\infty d\omega_K\left[\hat{a}_{\omega_K}\left(\alpha_{\omega_K \omega}^{\text{int}}\tilde{T}_H^{\text{int}}+\beta_{\omega_K \omega}^{\text{ext}}R_H^{\text{int} *}\right)\right. \quad \quad \quad \quad \quad \quad \nonumber \\
&\quad \quad\left.+\hat{a}_{\omega_K}^\dagger \left(\beta_{\omega_K \omega}^{\text{int} *}\tilde{T}_H^{\text{int}}+\alpha_{\omega_K \omega}^{\text{ext} *}R_H^{\text{int} *}\right)\right]+^{\text{in}}\!\hat{a}_I^{\text{ext}\dagger}R_I^{\text{int}*} \;,\quad \label{Eq:AnihhilationOperator2}\\
^{\text{out}}\!\hat{a}_{\text{ds}}^{\text{int}}=(\hat{\phi}^{(2)},\  ^{\text{out}}\!f_{\text{ds}}^{\text{int}})=&\int_0^\infty d\omega_K\left[\hat{a}_{\omega_K}\left(\alpha_{\omega_K \omega}^{\text{ext}}T_H^{\text{int}}+\beta_{\omega_K \omega}^{\text{int}}\tilde{R}_H^{\text{int} *}\right)\right.\quad \quad \quad \quad \quad \quad \nonumber \\
&\quad \quad\left.+\hat{a}_{\omega_K}^\dagger \left(\beta_{\omega_K \omega}^{\text{ext} *}T_H^{\text{int}}+\alpha_{\omega_K \omega}^{\text{int} *}\tilde{R}_H^{\text{int} *}\right)\right]+^{\text{in}}\!\hat{a}_I^{\text{ext}}T_I^{\text{int}}\;.\quad \label{Eq:AnihhilationOperator3}
\end{align}
\end{subequations}
 The  Bogoliubov coefficients  relating these annihilation operators are given by \footnote{These  have been calculated in Ref.~\cite{AndBal1240182013}, but the expressions there are missing a factor of $\kappa^{\pm\frac{i\omega}{\kappa}}$.}
\bes \bea
\alpha^{\text{ext}}_{\omega_K \omega}=\frac{1}{2\pi\kappa}\sqrt{\frac{\omega}{\omega_K}}\left(\frac{-i\omega_K}{\kappa}\right)^{\frac{i\omega}{\kappa}}\Gamma \left(\frac{-i\omega}{\kappa}\right)\ ,  \\
\beta^{\text{ext}}_{\omega_K \omega}=\frac{1}{2\pi\kappa}\sqrt{\frac{\omega}{\omega_K}}\left(\frac{-i\omega_K}{\kappa}\right)^{-\frac{i\omega}{\kappa}}\Gamma \left(\frac{i\omega}{\kappa}\right)\ ,  \\
\alpha^{\text{int}}_{\omega_K \omega}=\frac{1}{2\pi\kappa}\sqrt{\frac{\omega}{\omega_K}}\left(\frac{i\omega_K}{\kappa}\right)^{-\frac{i\omega}{\kappa}}\Gamma \left(\frac{i\omega}{\kappa}\right)\ ,  \\
\beta^{\text{int}}_{\omega_K \omega}=\frac{1}{2\pi\kappa}\sqrt{\frac{\omega}{\omega_K}}\left(\frac{i\omega_K}{\kappa}\right)^{\frac{i\omega}{\kappa}}\Gamma \left(\frac{-i\omega}{\kappa}\right).\label{Eq:BogluibovCoef}
\eea \ees
\subsection{The Hawking-partner correlator}
The main peak which was found in the density-density correlation function for the experimental results \cite{Ste2016959,Nature.569.s41586} is composed of modes which are traveling upstream toward $I_+$. The main contribution to these modes can be understood as arising from a Hawking particle in the exterior and its partner in the interior. The peak was Fourier decomposed to show the resulting correlation spectrum
in~\cite{Nature.569.s41586}.
 It was shown in \cite{PhysRevD.92.024043} that this correlation can be described by the quantity $S_0^2\left|\left<(\ ^{\text{out}}\hat{a}_{\text{up}}^{\text{ext}})( \ ^{\text{out}}\hat{a}_{\text{up}}^{\text{int}} )\right>\right|^2$, where $S_0$ is defined as the zero temperature static structure factor in Ref. \cite{Ste2016959}(see also~\cite{PhysRevD.92.024043}).  For relatively low momenta, which we will consider in our calculations, it is a good approximation to replace $S_0^2$ with $A\omega^2$, where $A$ is a constant that we will set to one.\footnote{This approximation for $S_0$ can be derived using results in~\cite{StrPit2003BEC}.}   For the other factor, we find
\be
\left|\left<U|(\ ^{\text{out}}\hat{a}_{\text{up}}^{\text{ext}})( \ ^{\text{out}}\hat{a}_{\text{up}}^{\text{int}} )|U\right>\right|^2=\left|\frac{e^{\frac{\pi\omega}{\kappa}}}{e^{\frac{2\pi\omega}{\kappa}}-1}(T_H^{\text{ext}}\tilde{T}_H^{\text{int}}+T_H^{\text{ext}}R_H^{\text{int} *}e^{\frac{\pi\omega}{\kappa}})+R_I^{\text{ext}}R_I^{\text{int} *}\right|^2, \label{Eq:HPCorrelatorGeneric}
\ee
where we have written the general expression in terms of scattering coefficients. We call $\left|\left<U|(\ ^{\text{out}}\hat{a}_{\text{up}}^{\text{ext}})( \ ^{\text{out}}\hat{a}_{\text{up}}^{\text{int}} )|U\right>\right|$ the HP correlator.

In the case where there is no effective potential and thus no scattering, $\tilde T_H^{\text{int}}=T_H^{\text{ext}}=1$,  $R_I^{\text{ext}}=R_I^{\text{int}}=0$, and Eq. \eqref{Eq:HPCorrelatorGeneric} reduces to
\be
\left|\left<U|(\ ^{\text{out}}\hat{a}_{\text{up}}^{\text{ext}})( \ ^{\text{out}}\hat{a}_{\text{up}}^{\text{int}} )|U\right>\right|^2=\left|\frac{e^{\frac{\pi\omega}{\kappa}}}{e^{\frac{2\pi\omega}{\kappa}}-1}\right|^2.
\ee
In this case the upstream modes which are thermally populated on the past horizon, simply travel toward $I_+$.  This expression only depends on the surface gravity and thus the Hawking temperature, $T_H=\frac{\kappa}{2\pi}$. However, if the effective potential is included,  the resultant quantities are also dependent on the details of the sound speed and velocity profiles
away from the horizon.

\subsection{Interior upstream and downstream phonon numbers \label{Sec:IntParticleNumber}}

In  \cite{PhysRevD.100.105021}, a new feature was found that is related to the interior DPN, $n_{\text{ds}}^{\text{int}}$, and the UPN,  $n_{\text{up}}^{\text{int}}$.  UPN refers to the number of phonons which are traveling upstream in the frame of the fluid in the interior, being dragged away from the horizon in the lab frame and arriving at $I_{\text{up}}^{\text{int}}$, while DPN  refers to phonons which are moving downstream and arriving at $I_{\text{ds}}^{\text{int}}$.
 The DPN and UPN expressed in terms of the creation and annihilation operators for the {\it out} modes are
 \be
n_{\text{up}}^{\text{int}}= \left<U|^{\text{out}}\hat{a}_{\text{up}}^{\text{int}\dagger }\ ^{\text{out}}\hat{a}_{\text{up}}^{\text{int}\ }|U\right> \quad \text{and}\quad  n_{\text{ds}}^{\text{int}}= \left<U|^{\text{out}}\hat{a}_{\text{ds}}^{\text{int}\dagger}\ ^{\text{out}}\hat{a}_{\text{ds}}^{\text{int}\ }|U\right>.\label{Eq:PhononNumbers}
 \ee
Solving  for $\left|n_{\text{up}}^{\text{int}}\right|^2$ using the definition for the annihilation operator in Eq.~\eqref{Eq:AnihhilationOperator2}, one finds
\be
\left|n_{\text{up}}^{\text{int}}\right|^2=\left(\frac{1}{e^{\frac{2\pi\omega}{\kappa}}-1}\left|\tilde{T}_H^{\text{int}}+R_H^{\text{int} *}e^{\frac{\pi\omega}{\kappa}}\right|^2+\left|R_I^{\text{int} *}\right|^2\right)^2. \label{Eq:UpParticleNumberGeneric}
\ee
 In the case where $V_{\text{eff}}=0$,  $\tilde{T}_H^{\text{int}}=1$, and $R_H^{\text{int} *}=R_I^{\text{int} *}=0$,
 \be
 \left|n_{\text{up}}^{\text{int}}\right|^2=\left(\frac{ 1}{e^{\frac{2\pi \omega}{ \kappa}}-1}\right)^2.\label{Eq:UPNNoScattering}
 \ee
Thus, there is a thermal distribution of phonons.

For the DPN, one finds
 \be
\left|n_{\text{ds}}^{\text{int}}\right|^2=\left(\frac{1}{e^{\frac{2\pi\omega}{\kappa}}-1}\left|{T}_H^{\text{int}}+\tilde{R}_H^{\text{int} *}e^{\frac{\pi\omega}{\kappa}}\right|^2\right)^2. \label{Eq:DsParticleNumberGeneric}
\ee
${T}_H^{\text{int}}$ is associated with a mode in the exterior that is positive frequency on $H_-$ and is partially scattered into the interior. Thus, in the absence of a potential   ${T}_H^{\text{int}}=0 $ and  $\tilde{R}_H^{\text{int} *}=0$, there is no particle production for these modes in the interior of the  analog black hole.

\section{ HP correlator with an Effective Potential  \label{Sec:HPCorrelatorWithPotential}}

We now apply this general formalism to the three models previously mentioned.

\subsection{Two-delta function potential}
The first model we
 consider  was discussed in \cite{PhysRevD.100.105021} where  $V_\text{eff}$ was approximated by two Dirac delta functions, one in region $r$ and one in region $l$. We refer to this model as the two-delta function potential model.  The effective potential is
\begin{align}
V_{\text{eff}}= \left\{
        \begin{array}{ll}
            V_{\text{int}}\delta (x^*-x^*_{\text{int}})\ , & \quad x < 0\ , \\
           V_{\text{ext}}\delta (x^*-x^*_{\text{ext}})\ , & \quad x > 0\ .
        \end{array}
    \right.   \label{2delta}
\end{align}
We review the resulting solution for the   $^{\rm in}f^{\text{ext}}_I$ modes for the entire spacetime. In the exterior, it is given by
 \bea
  ^{\rm in}f^{\text{ext}}_I &=& \frac{e^{-i\omega t}}{\sqrt{4\pi\omega}}\left[ e^{-i \w x^*} + R^{\text{ext}}_I e^{i \w x^*}\right]  \,, \qquad x^{*} > x^{*}_{\text{ext}} = 0  \;, \nonumber \\
                        &=&\frac{e^{-i\omega t}}{\sqrt{4\pi\omega}} T^{\text{ext}}_I e^{-i \w x^*}  \,, \qquad x^{*} < x^{*}_{\text{ext}} = 0  \;,
\eea
where $R$ and $T$ will refer to scattering coefficients throughout.    In the interior,
 \bea
  ^{\rm in}f^{\text{ext}}_I &=&\frac{e^{-i\omega t}}{\sqrt{4\pi\omega}} T^{\text{ext}}_I e^{-i \w x^*} \,, \qquad x^{*} < x^{*}_{\text{int}} = 0  \;, \nonumber \\
                        &=&  \frac{e^{-i\omega t}}{\sqrt{4\pi\omega}}\left[ T^{int}_I e^{-i \w x^*} +  R^{int}_I e^{i \w x^*} \right] \,, \qquad x^{*} > x^{*}_{\text{int}}= 0  \;.
 \eea
The asymptotic form of the $ ^{\text{in}}\!f_I^{\text{ext}}$ mode as $x\to +\infty$ is
\be\label{ciao}
^{\text{in}}\!f_I^{\text{ext}}\to\frac{e^{-i\omega t}}{\sqrt{4\pi \omega}}\left[ e^{-i\omega x^*}+R_I^{\text{ext}}e^{i\omega x^*}\right]
\ee
 and for $x\to -\infty$ it has the form
 \be
^{\text{in}}\!f_I^{\text{int}}\to\frac{e^{-i\omega t}}{\sqrt{4\pi \omega}}\left[ T_I^{\text{int}}e^{-i\omega x^*}+R_I^{\text{int}}e^{i\omega x^*}\right].
\ee

Similarly, the modes which originate on the exterior past horizon have  the following asymptotic form for  $x\to +\infty$:
\be \label{ciao2}
^{\text{in}}\!f_H^{\text{ext}}\to\frac{e^{-i\omega t}}{\sqrt{4\pi \omega}}T_H^{\text{ext}}e^{i\omega x^*}.
\ee
 The form  for  $x\to -\infty$ is
 \be
^{\text{in}}\!f_H^{\text{ext}}\to\frac{e^{-i\omega t}}{\sqrt{4\pi \omega}}\left[ R_H^{\text{int}}e^{-i\omega x^*}+T_H^{\text{int}}e^{i\omega x^*}\right]\label{Eq:InMode2c}.
\ee
Finally, the modes which originate on the past horizon in the interior have  the  form for $x\to -\infty$,
\be
^{\text{in}}\!f_H^{\text{int}}\to\frac{e^{i\omega t}}{\sqrt{4\pi \omega}}\left[ \tilde{T}_H^{\text{int}}e^{-i\omega x^*}+\tilde{R}_H^{\text{int}}e^{i\omega x^*}\right]\label{Eq:InMode3c}\ .
\ee

The transmission and reflection coefficients have been computed in~\cite{PhysRevD.100.105021}. They are found by enforcing continuity of the radial mode functions at the locations of the delta function potentials and imposing the usual jump conditions on the first derivatives of the radial mode functions at those locations.  The jump conditions are obtained by integrating the mode equation \eqref{Eq:KGEMMSFNCVofX} around the delta function potential over an interval $\left[-\epsilon, \epsilon \right]$ in the limit $\epsilon \to 0 $.
  The resulting scattering coefficients are
\bea
T_I^{\text{ext}}=\frac{\frac{2i\omega}{V_{\text{ext}}}}{\frac{2i\omega}{V_{\text{ext}}}-1}\ , &\quad \quad  T_H^{\text{ext}}=\frac{1}{1-\frac{V_{\text{ext}}}{2i\omega}}\ ,\nonumber \\
R_I^{\text{ext}}=\frac{1}{\frac{2i\omega}{V_{\text{ext}}}-1}\ , & \quad \quad R_H^{\text{ext}}=\frac{\frac{V_{\text{ext}}}{2i\omega}}{1-\frac{V_{\text{ext}}}{2i\omega}}\ ,\nonumber\\
R_I^{\text{int}}=\frac{{V_{\text{int}}}}{{2i\omega}}T_I^{\text{ext}}\ , & \quad \quad T_H^{\text{int}}=\left(1-\frac{V_{\text{int}}}{2i\omega}\right)R_H^{\text{ext}}\nonumber\\
T_I^{\text{int}}=\left(1-\frac{V_{\text{int}}}{2i\omega}\right)T_I^{\text{ext}}\ ,& \quad \quad R_H^{\text{int}}=\frac{{V_{\text{int}}}}{{2i\omega}}R_H^{\text{ext}}
\nonumber\\
\tilde{T}_H^{\text{int}}=1-\frac{V_{\text{int}}}{2i\omega}\ ,& \quad \quad \tilde{R}_H^{\text{int}}=\frac{{V_{\text{int}}}}{{2i\omega}}.
 \eea

Using these scattering coefficients in Eq.~\eqref{Eq:HPCorrelatorGeneric} gives
\begin{align}
\left|\left<U|(^{\text{out}}\hat{a}_{\text{up}}^{\text{ext}})\right.\right.  &\left.\left. ( ^{\text{out}}\hat{a}_{\text{up}}^{\text{int}} )|U\right>\right|^2\nonumber \\
=&\left|\frac{e^{\frac{\pi\omega}{\kappa}}}{e^{\frac{2\pi\omega}{\kappa}}-1}\left(\frac{2i\omega-V_{\text{int}}}{2i\omega-V_{\text{ext}}}-\frac{V_{\text{int}}V_{\text{ext}}}{4\omega^2+V_{\text{ext}}^2}e^{\frac{\pi \omega}{\kappa}}\right)+\frac{V_{\text{int}}}{V_{\text{ext}}+\frac{4\omega^2}{V_{\text{ext}}}}\right|^2
\end{align}

In the two-delta function potential model, the HP correlator is finite as $\omega\to 0$ whereas in the case without scattering it diverges in this limit.
This can be seen in the quantity $\omega^2 \left|\left<U|(^{\text{out}}\hat{a}_{\text{up}}^{\text{ext}})(^{\text{out}}\hat{a}_{\text{up}}^{\text{int}} )|U\right>\right|^2$ which is plotted in Fig.~2 for both the two-delta function potential and the case with no scattering. The ratio of the two cases is also shown.

\begin{figure}[h!]
\includegraphics[width=2.9in]{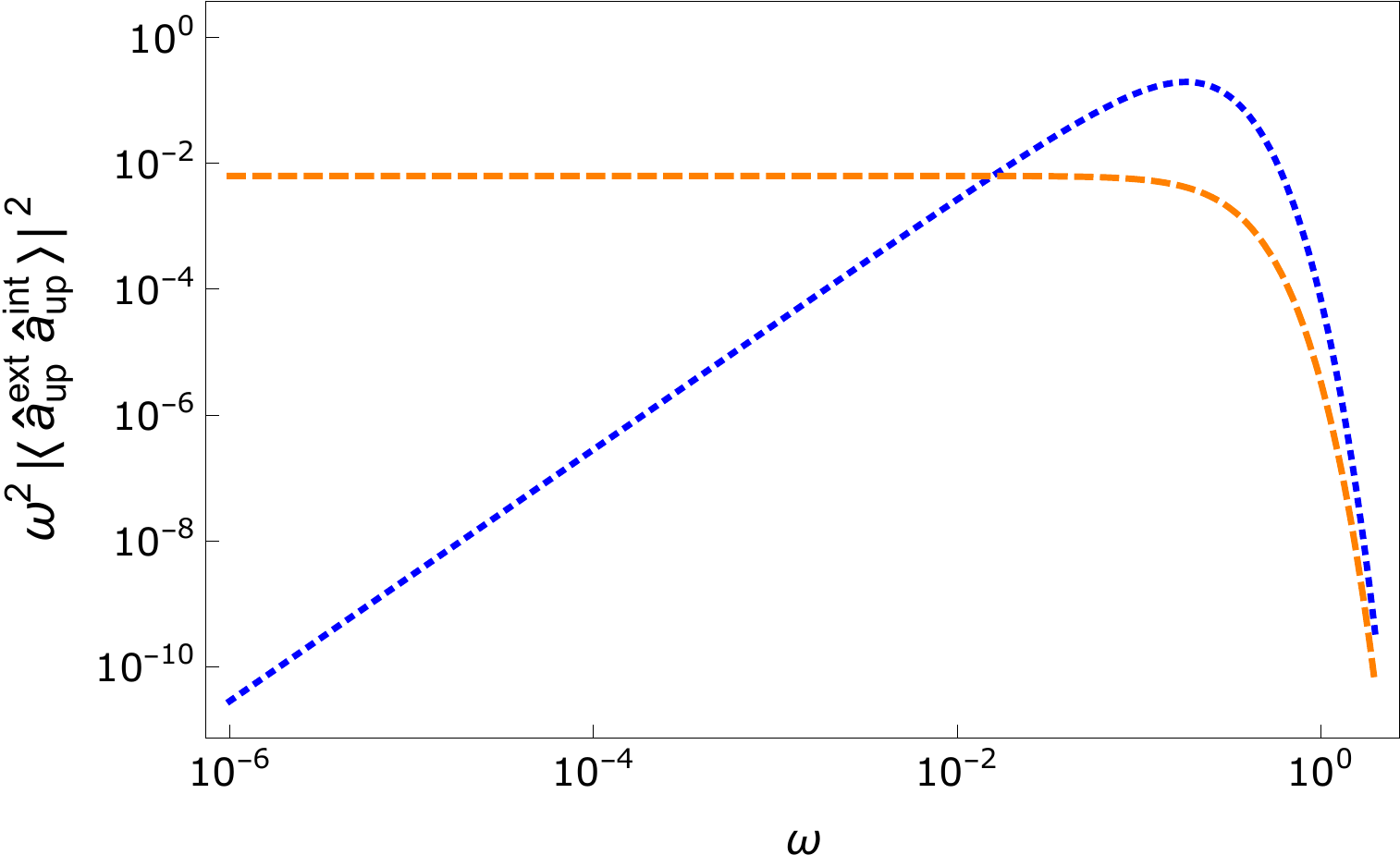}\hspace{.15in}
\includegraphics[width=2.7in]{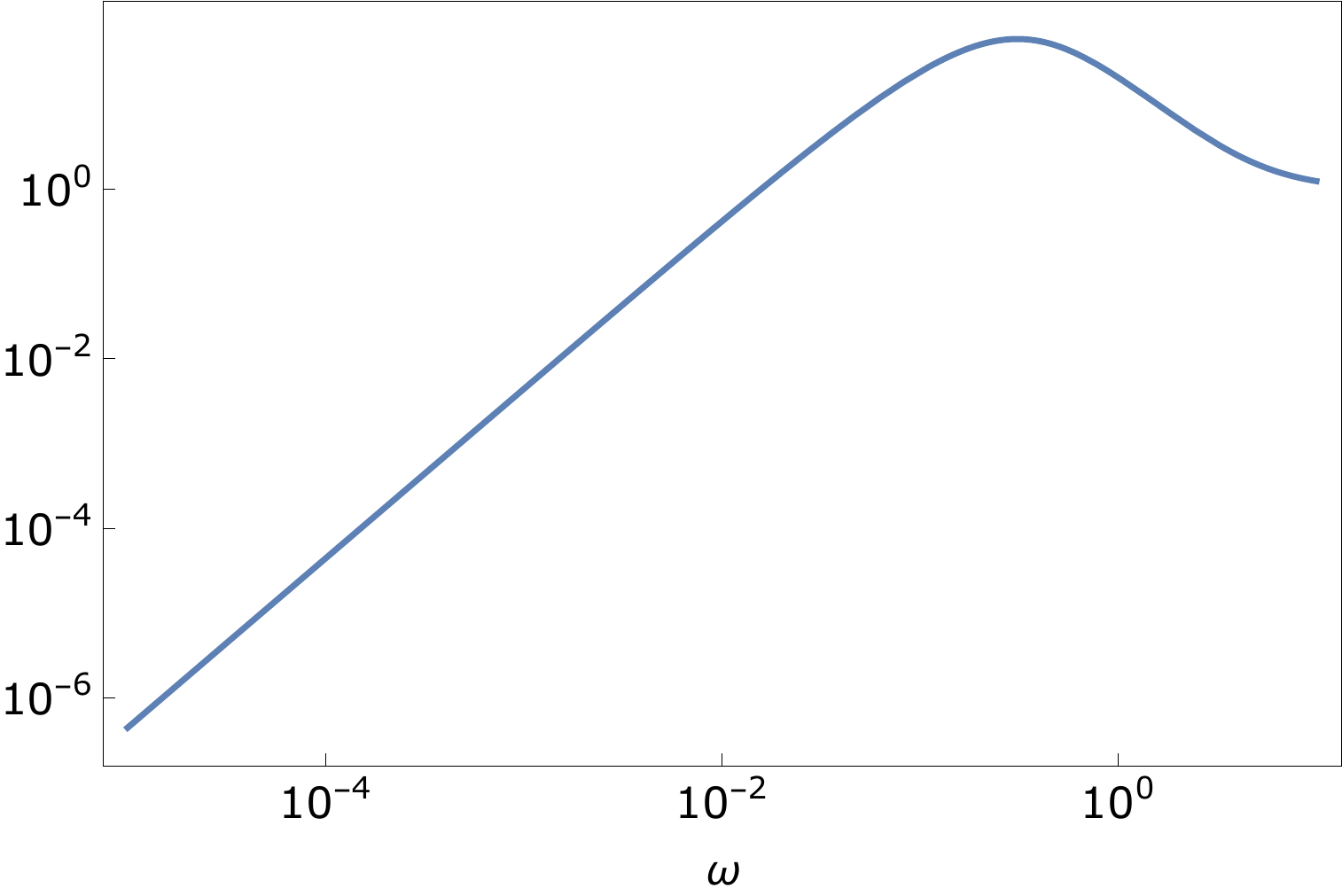}
\caption{\label{Fig:THComparisonsdelta} The square of the product of the Hawking-partner correlator and the frequency for the two-delta function potential model (blue dotted) and the case with $V_{\text{eff}}=0$ (orange dashed) is shown in the left panel. Here $\left|\left<^{\text{out}}\!\hat{a}_{\text{up}}^{\text{ext}}\ ^{\text{out}}\!\hat{a}_{\text{up}}^{\text{int}}\right>\right|$ is the Hawking-partner correlator.  In the right panel,  the ratio of the two curves on the left is shown.  }
\end{figure}

\clearpage

\subsection{Constant flow velocity model}
\label{Sec:ConstantFlowModel}
 We
  next consider a model which has been studied from both the condensed matter perspective \cite{PhysRevA.80.043603} and the quantum field theory in curved space perspective \cite{AndBal1240182013} and shows good agreement between the two.
The profile has a varying  sound speed, but the flow velocity is held constant, and thus due to mass continuity, the density is also constant.  Such a profile is theoretically possible if an external potential is adjusted to keep the density constant while the coupling constant, $g$, which is related to the s-wave scattering length, is varied via a Feshbach resonance\cite{StrPit2003BEC} allowing the speed of sound,  $c=\sqrt{\frac{gn}{m}}$ to vary.  \\
 The sound speed profile used in~\cite{CarFag2008103001,AndBal1240182013} is\footnote{In~\cite{CarFag2008103001} this profile was used with $b = 0$.}
\bea  c(x)&=&\sqrt{c_{\text{int}}^2 + \frac{1}{2}(c_{\text{ext}}^2-c_{\text{int}}^2)\left[1+\frac{2}{\pi}\tan^{-1}\left(\frac{x+b}{\sigma_{v}}\right)\right]}  \;, \nonumber \\
      b &=& \sigma_{v} \tan \left[\frac{\pi}{c_{\text{ext}}^2-c_{\text{int}}^2} \left( v_0^2 - \frac{1}{2} (c_{\text{ext}}^2+c_{\text{int}}^2)\right) \right] \;, \label{c-used2}
 \eea
 where $b$ is defined so that the horizon occurs at $x=0$.  This sound speed approaches a constant, $c_{\text{int}}$ in the interior, as $x\to -\infty$ and in the exterior approaches the constant $c_{\text{ext}}$ as $x\to \infty$. The flow velocity is $\vec{v}=-v_0\hat{x}$, where $v_0>0$ is constant. The term $\sigma_{v}$ is related to the width of the profile. We use $c_{\text{int}}=1/2$, $c_{\text{ext}}=1$, $v_0=3/4$, and $\sigma_{v}=8$ which are the values used for some of the numerical calculations in~\cite{AndBal1240182013}.

The scattering coefficients are calculated numerically for each value of $\omega$ and then used in Eq. \eqref{Eq:HPCorrelatorGeneric}. Unlike the two-delta function potential case, the reflection coefficient in the exterior  does not approach one for low frequencies; thus, the HP correlator is infrared divergent as it is when the effective potential is ignored.

\begin{figure}[h!]
\includegraphics[width=2.9in]{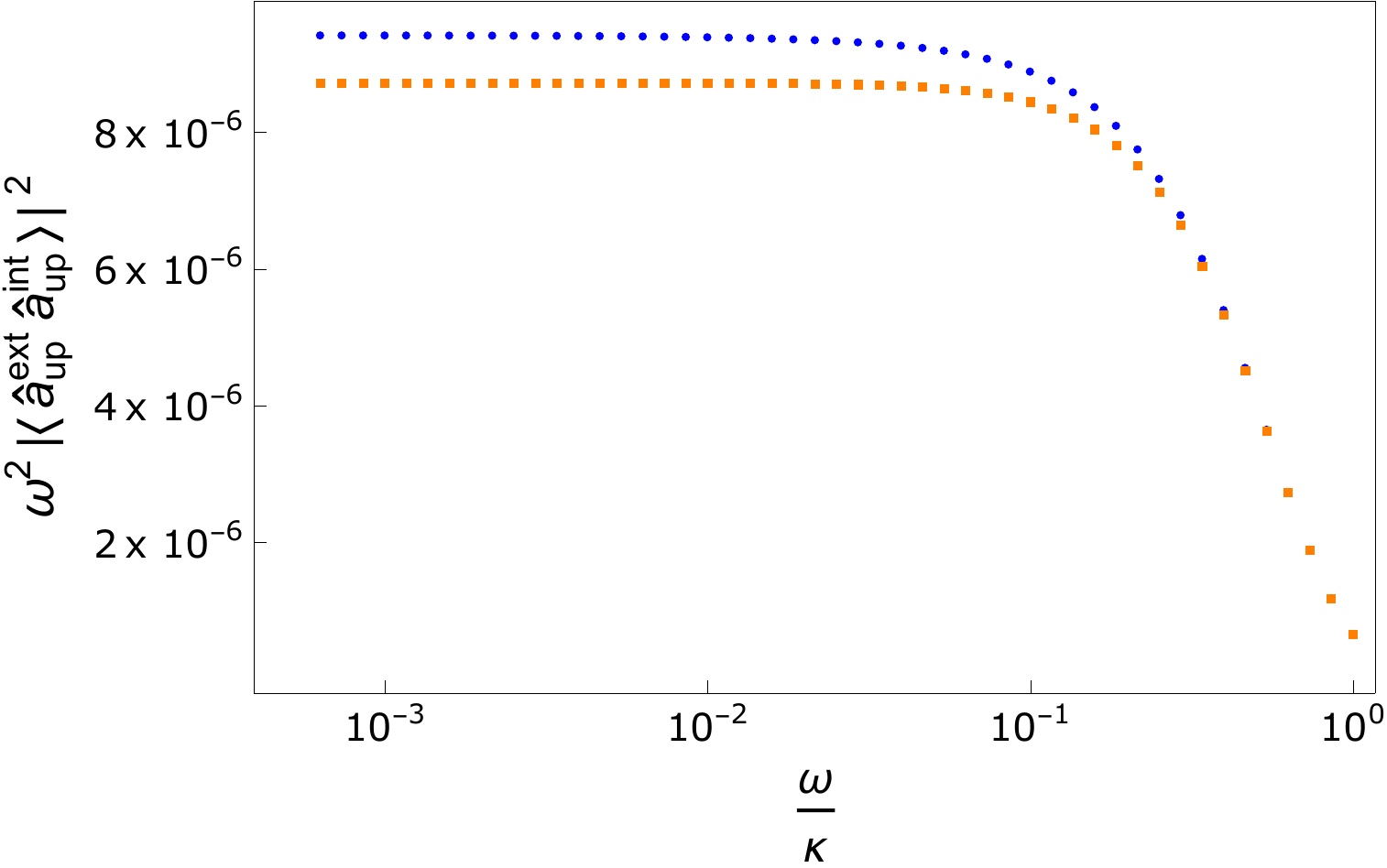}\hspace{.15in}
\includegraphics[width=2.6in]{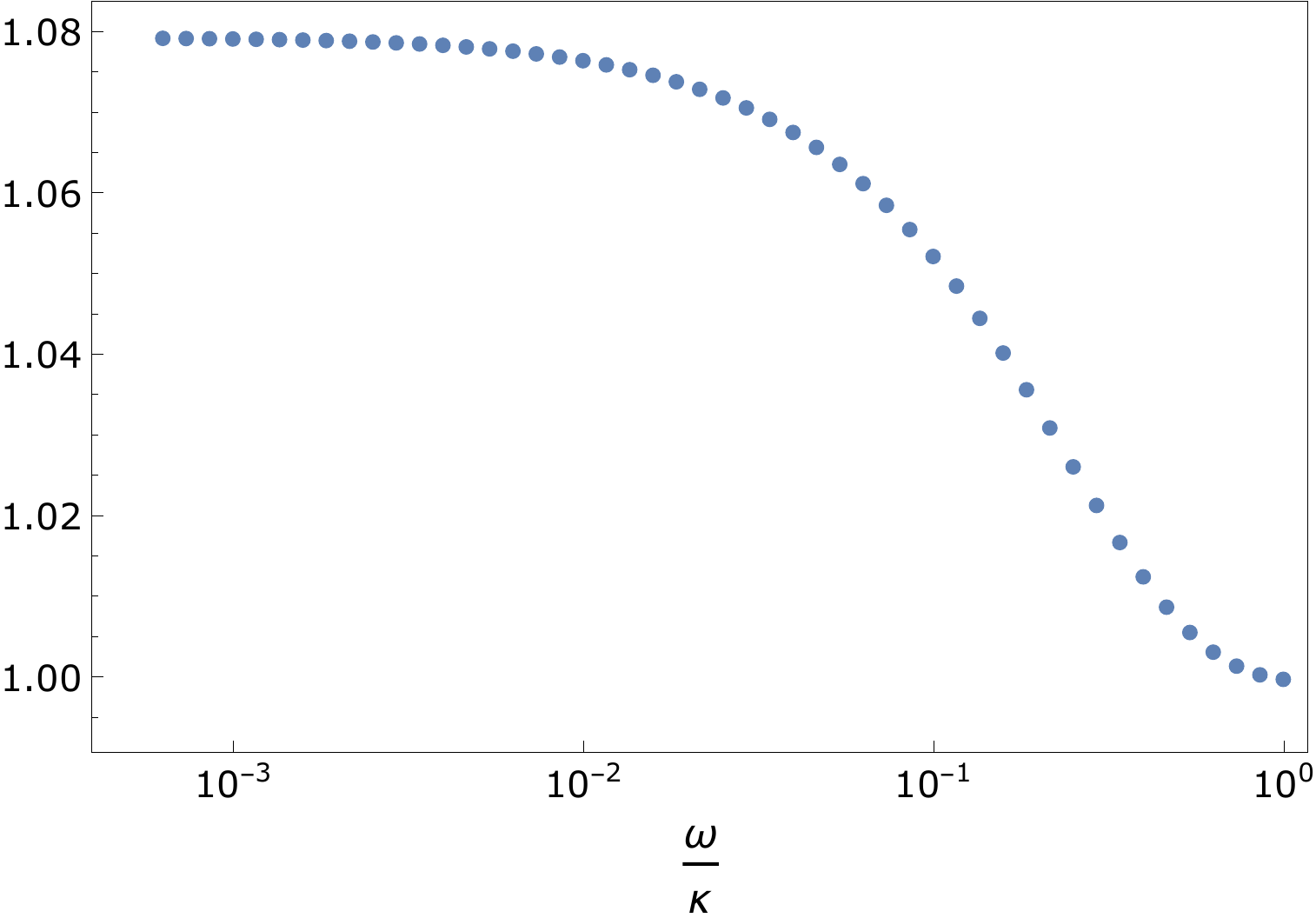}
\caption{\label{Fig:2013HPCorrelatorComparisons}  The square of the product of the Hawking-partner correlator and the frequency for the constant flow velocity model (blue dots) and the case with $V_{\text{eff}}=0$ (orange squares) is shown in the left panel. Here $\left|\left<^{\text{out}}\!\hat{a}_{\text{up}}^{\text{ext}}\ ^{\text{out}}\!\hat{a}_{\text{up}}^{\text{int}}\right>\right|$ is the Hawking-partner correlator.  In the right panel  the ratio of the two curves on the left is shown.  }
\end{figure}

    The results are shown in Fig. \ref{Fig:2013HPCorrelatorComparisons} where the quantity   $\omega^2\left|\left<^{\text{out}}\hat{a}_{\text{up}}^{\text{ext}} \; ^{\text{out}}\hat{a}_{\text{up}}^{\text{int}}\right>\right|^2$ is plotted both when $V_{\text{eff}}$ is included in the calculation and when $V_{\text{eff}}=0$. The inclusion of the effective potential increases  the value of the HP correlator  throughout the frequency  range of the plot.  A ratio of the two cases is also shown. In the low frequency regime, the HP correlator is observed to be approximately $ 8\% $ larger than its value when $V_{\text{eff}}=0$. This inevitably will affect the main peak.

\subsection{The waterfall model}
A model, which more closely resembles the experiments of \cite{Ste2016959} and \cite{Nature.569.s41586}, but which still has some significant differences, has been studied in  \cite{2016PhRvD94h4027M}. This model, often called the waterfall model, is based on an analytic solution to the Gross-Pitaevskii equation when an external step function potential is applied.  The resulting  density profile can be written as
\be
{n}(x)=\left\{
\begin{array}{cc}
{n}_{-}\left[M_-+(1-M_-)/(\cosh(\sigma( x+x_0)))^2\right]& x\geq x_0 \\
{n}_{-}& x\leq x_0.
\end{array}
\right.
\ee
where we have shifted the profile by $x_0\approx9.6\times10^{-7}$ so that the horizon is at $x=0$. The subscript $-$ indicates the asymptotic value as $x\to -\infty $. The Mach number $M(x)\equiv c(x)/v_0(x)$ is used to characterize the flow, and its asymptotic value ${M_-=c_{-} /v_{0-} }$ gives insight into the strength of the ``waterfall''.  The width of the profile is modified by ${\sigma=(\sqrt{M_--1})/\xi}$ with $\xi$ the healing length. In this profile the flow velocity,  sound speed,
and density all vary along the flow. The flow velocity is  $\vec{v}=-v_0(x)\hat{x}$, with $v_0(x)>0$. It is simple to show that the continuity equation leads to $v_0\propto \frac{1}{n}$ (see e.g.~\cite{lamb1916hydrodynamics}).   \\

\begin{figure}[h!]
\centering
	\includegraphics[width=5.in, height=2.2in]{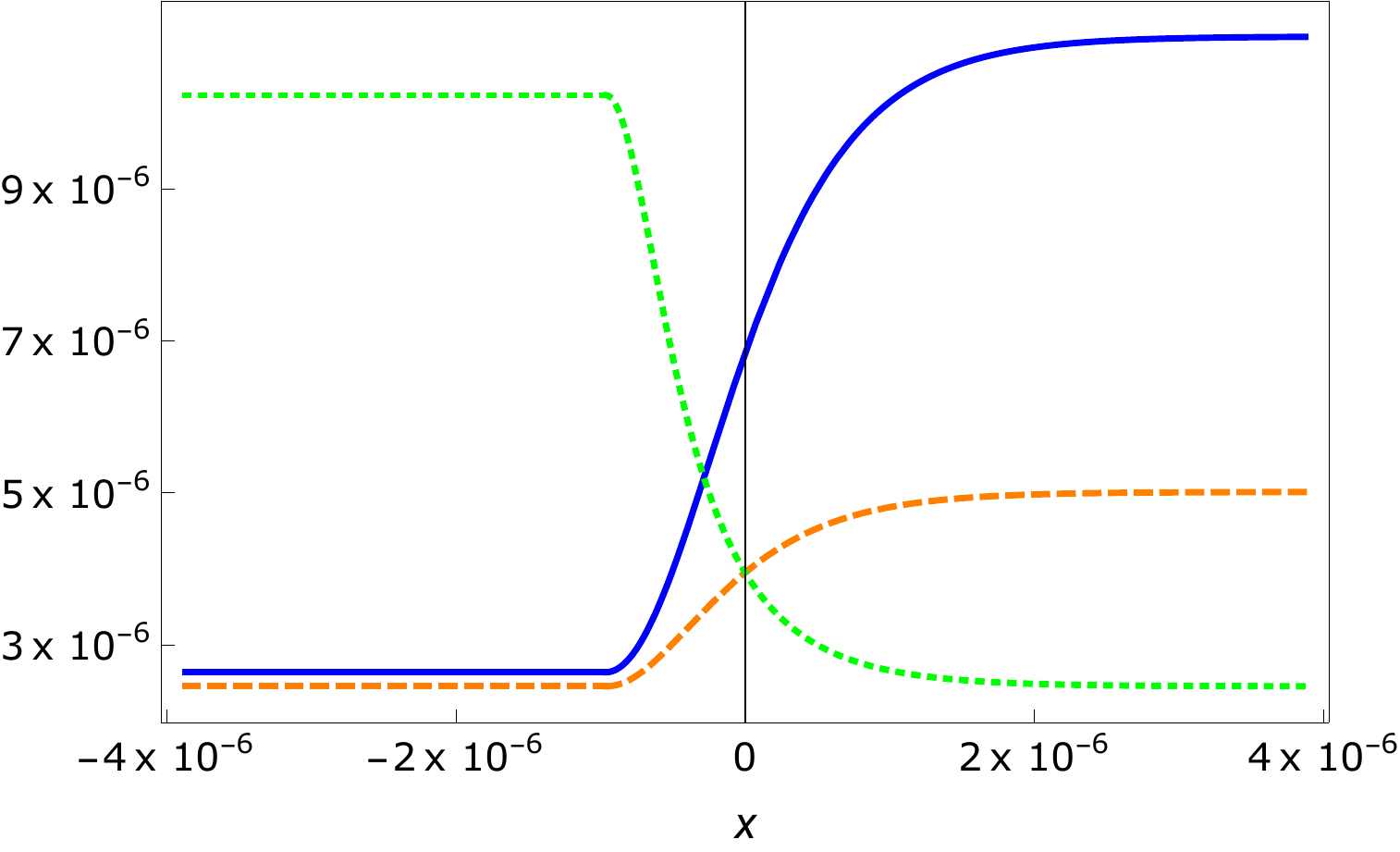}	
	\caption{\label{Fig:THComparisons} Various profiles are shown for the waterfall model with Mach number $M_- = 4.1$.  The solid (blue) curve corresponds to the density $n(x)$ multiplied by the factor $2.7 \times 10^{-11}$; the dashed (orange) curve corresponds to the sound speed $c(x)$, and the dotted (green) curve corresponds to the flow speed $|v(x)|$. }
\end{figure}

  The entire solution can be defined by a particular choice for $c_{-} $ and  $v_{-}$. Here we use values that loosely match the experiment described in \cite{Ste2016959} with  $v_-=1.02\times10^{-3}$ and $c_-=0.24\times10^{-3}$. The resulting density, sound speed, and flow velocity are plotted in Fig.~\ref{Fig:THComparisons}.

\begin{figure}[h!]
	\includegraphics[width=2.9in]{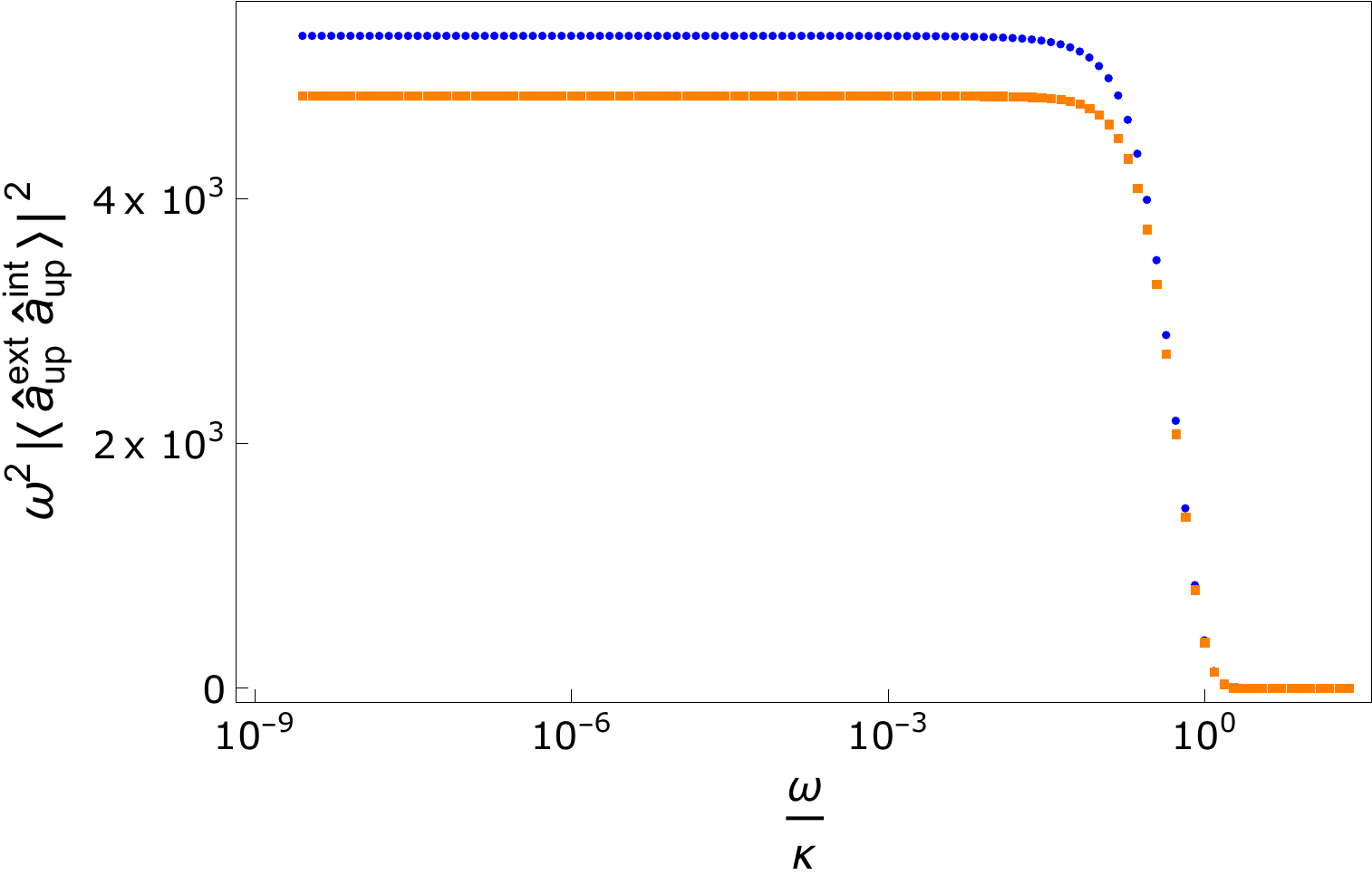}\hspace{.15in}
	\includegraphics[width=2.6in]{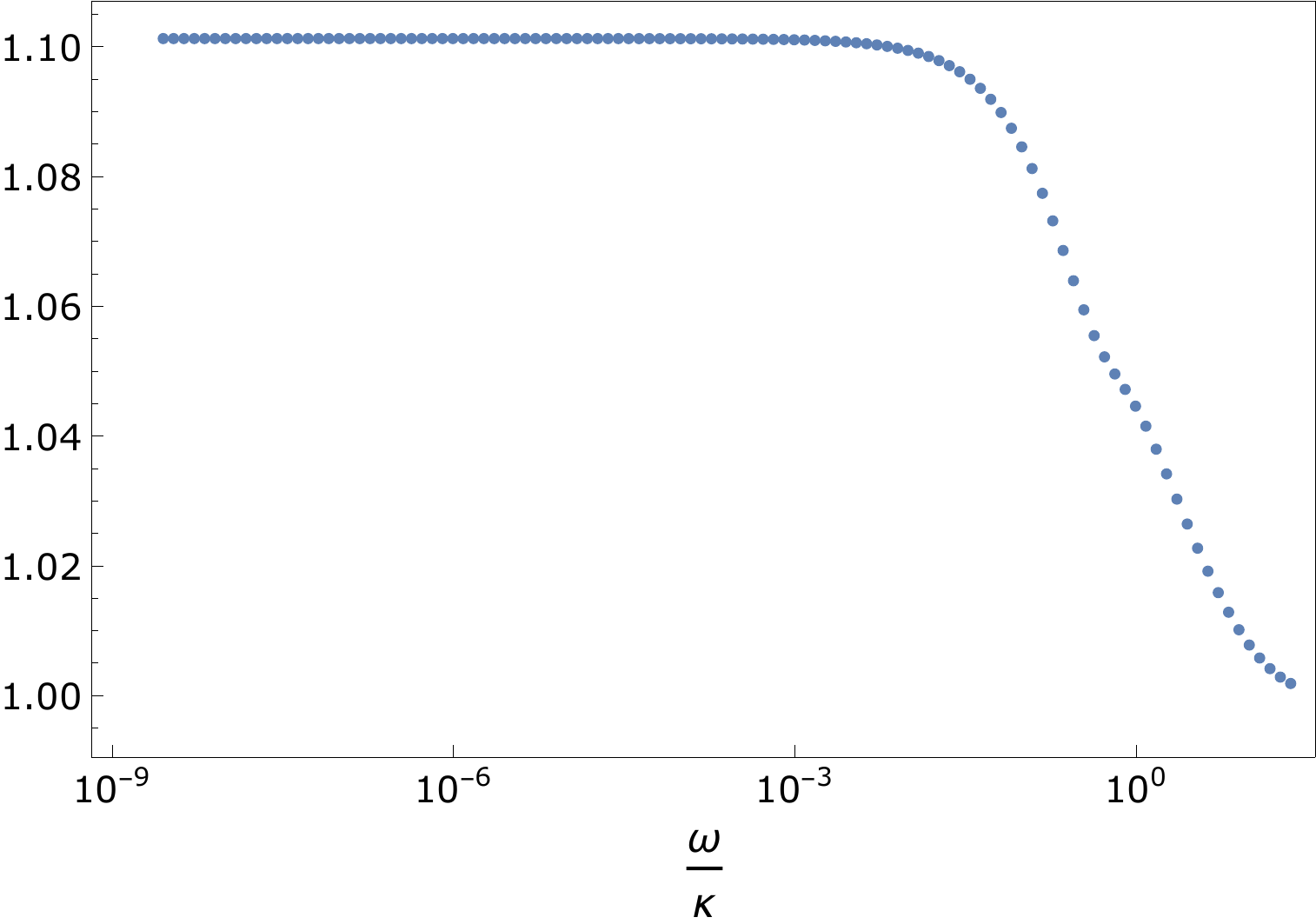}
	\caption{\label{Fig:WaterfallHPCorreleatorComparisons4} The square of the product of the Hawking-partner correlator and the frequency for the waterfall model (blue dots) and the case with $V_{\text{eff}}=0$ (orange squares) is shown in the left panel. Here $\left|\left<^{\text{out}}\!\hat{a}_{\text{up}}^{\text{ext}}\ ^{\text{out}}\!\hat{a}_{\text{up}}^{\text{int}}\right>\right|$ is the Hawking-partner correlator.  In the right panel,  the ratio of the two curves on the left is shown.}
\end{figure}

The result for the HP correlator is shown in Fig. \ref{Fig:WaterfallHPCorreleatorComparisons4}, where the quantity $\omega^2\left|\left<^{\text{out}}\hat{a}_{\text{up}}^{\text{ext}}\; ^{\text{out}}\hat{a}_{\text{up}}^{\text{int}}\right>\right|^2$ is plotted for $V_\text{eff}\neq 0$ and for $V_\text{eff}=0$. There is a significant difference between these two cases  throughout most of the frequency  range of the plot. The ratio of the two cases is also shown, and  there is an approximately $ 10\%$ increase in the low frequency values of the HP correlator when the effective potential is included in the calculation. This low frequency regime is especially important when considering the main peak in the density-density correlation function as both the width and magnitude of the peak are heavily dependent on the low frequency modes.

\section{Particle production in the interior\label{Sec:InteriorParticleProduction3Models}}
 The numbers of upstream and downstream phonons in the interior of a BEC analog black hole were computed in \cite{PhysRevD.100.105021} for the two-delta function potential model. We review these results and then calculate quantities related to the interior UPN and  DPN for the other two models.

\begin{figure}[h!]	
	\includegraphics[width=2.9in]{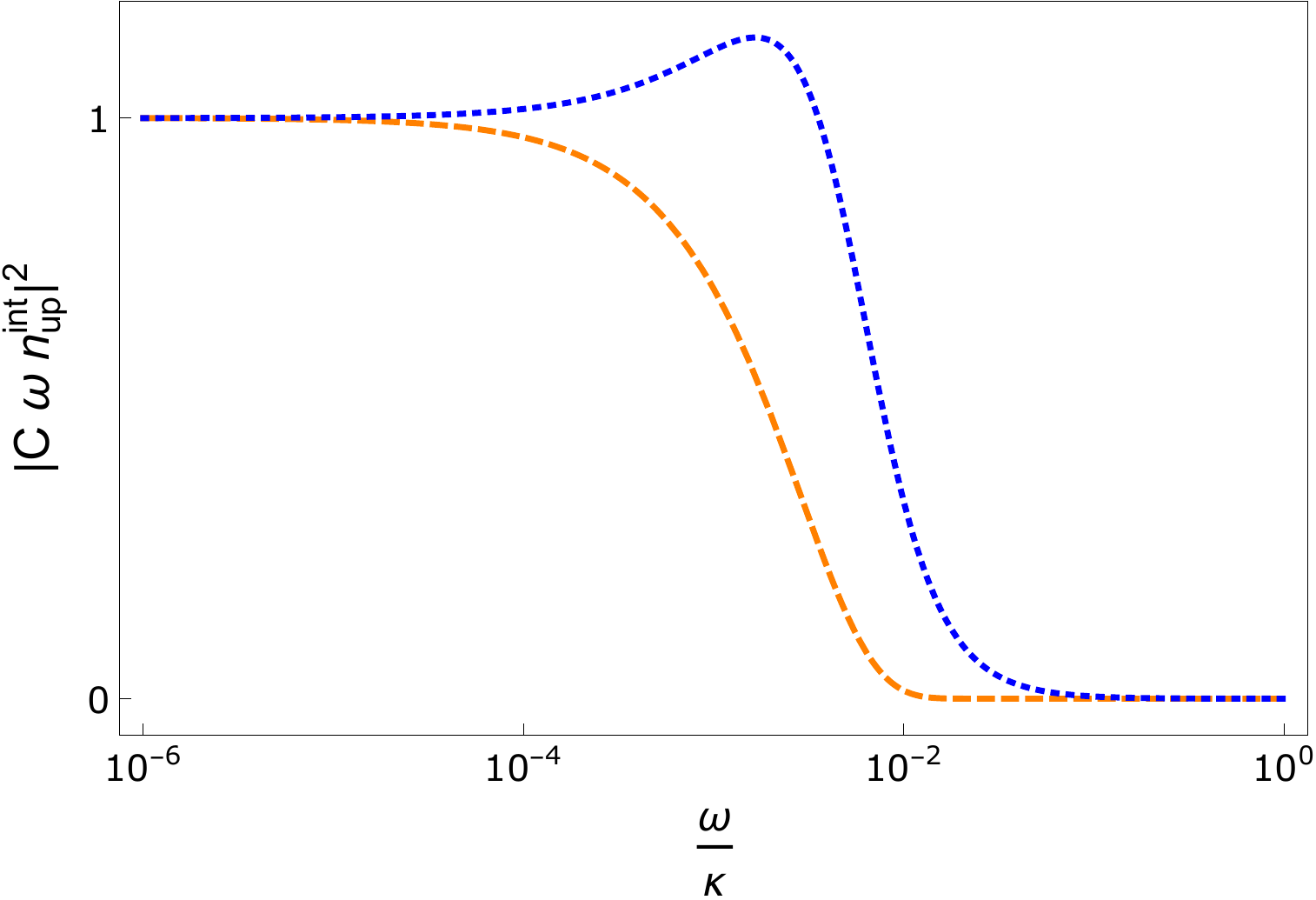}\hspace{.15in}
	\includegraphics[width=2.9in]{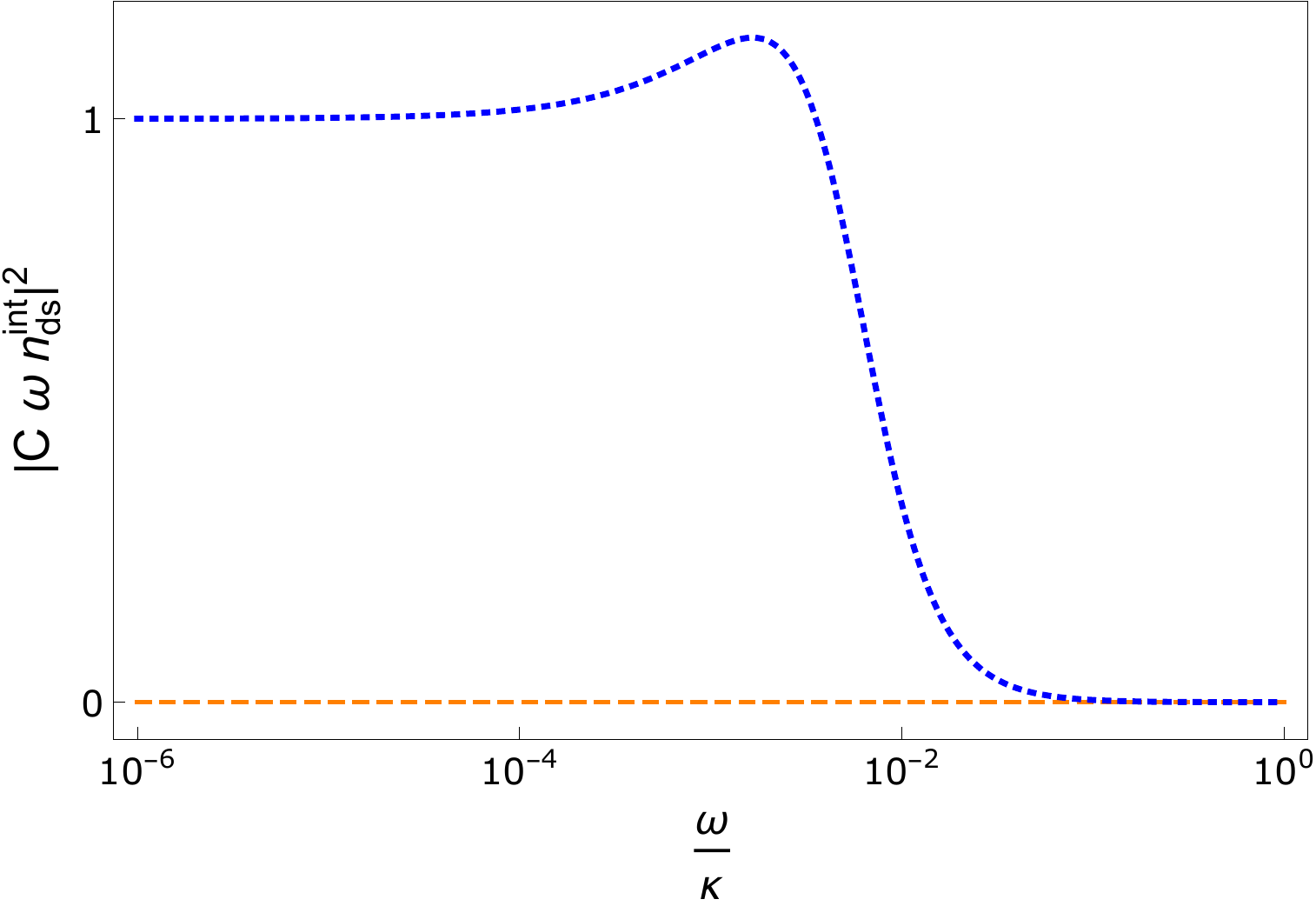}
	\caption{\label{Fig:WaterFallInteriorComparisons5}  In the left panel the quantity  $\left|C\ \omega \ n_{\text{up}}^{\text{int}}\right|^2$  is shown for the two-delta function potential model (blue dotted)  and for the case where there is no potential (orange dashed).  In the right panel the quantity  $\left|C\ \omega \ n_{\text{ds}}^{\text{int}}\right|^2$   is shown for the  two-delta function potential model(blue dotted)  and for the case where there is no potential (orange dashed). For the two-delta function potential model $V_{\text{int}}=-\kappa/100$ and  $V_{\text{ext}}=2\kappa/3$. $C$ is a  scaling factor whose value is chosen, where possible, for each curve so that
 $C\ \omega \ |n_{\text{up}}^{\text{int}}|=1$ or  $C\ \omega \ |n_{\text{ds}}^{\text{int}}|=1$ for ${\omega = 10^{-6}}$. }
\end{figure}

 In Sec.~\ref{Sec:IntParticleNumber} we have shown that if $V_{\text{eff}}=0$ the spectrum of $|\omega n_{\text{up}}^{\text{int}}|^2$  is based on a thermal distribution as seen in Eq.\eqref{Eq:UPNNoScattering} and $|\omega n_{\text{ds}}^{\text{int}}|^2=0$. For the two-delta function potential it was shown in \cite{PhysRevD.100.105021} that  $n_{\text{ds}}^{\text{int}}$ is nonzero  and that both $ n_{\text{up}}^{\text{int}}$ and $ n_{\text{ds}}^{\text{int}}$  are nonthermal in the left and right panels, respectively, of Fig.~\ref{Fig:WaterFallInteriorComparisons5}. In both plots, the spectrum for these quantities when $V_{\text{eff}}=0$ is  shown. Recalling that for $V_\text{eff}=0$, $|\omega n_{\text{up}}^{\text{int}}|^2$ has a thermal spectrum, it is clear that the spectrum when $V_\text{eff}\neq 0$  is nonthermal.

 Also visible in Fig. \ref{Fig:WaterFallInteriorComparisons5} is a peak. It was found in \cite{PhysRevD.100.105021} that this peak occurs when the magnitude of $V_\text{eff}$ is larger in the interior than it is in the exterior.

For the first model, the delta-function effective potential was introduced in an {\it ad hoc} way and
the asymmetry
in the overall potential profile is  thus not related to the sound speed or flow velocity of the model. In the other two models, the effective potential is derived from the
profiles for the sound speed and flow velocity according to~(\ref{effpot}).

\begin{figure}[h!]
	\includegraphics[width=5.5in]{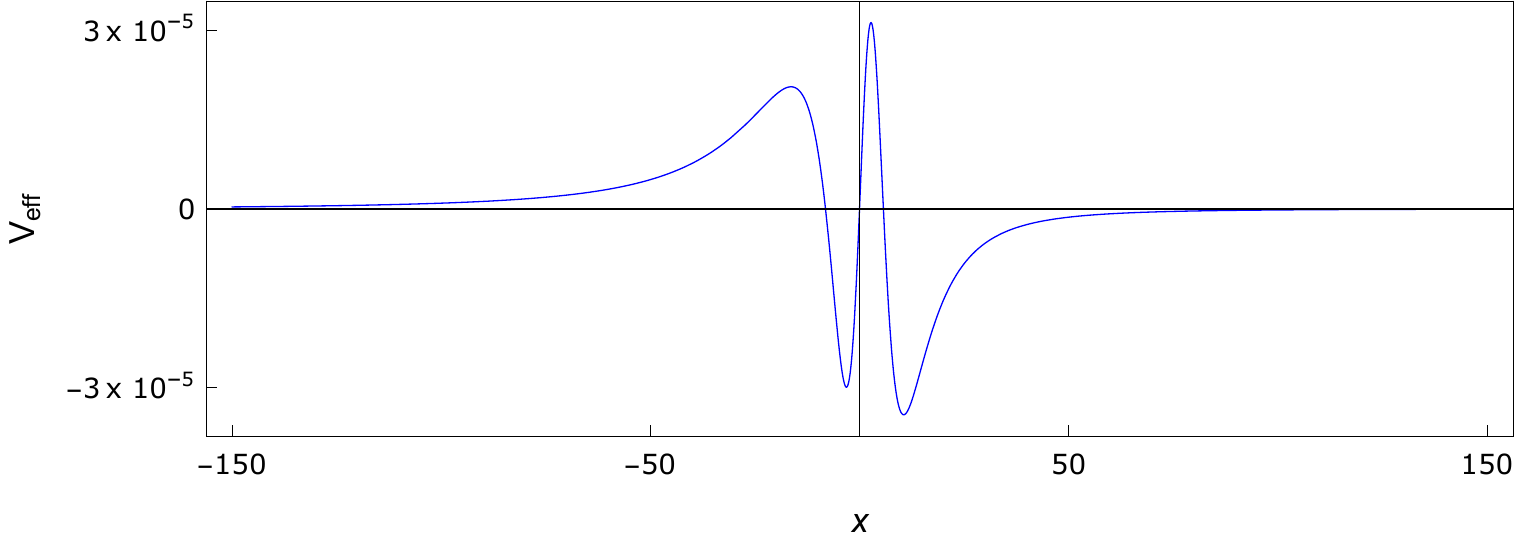}	\vspace{.1in}
	\includegraphics[width=2.85in]{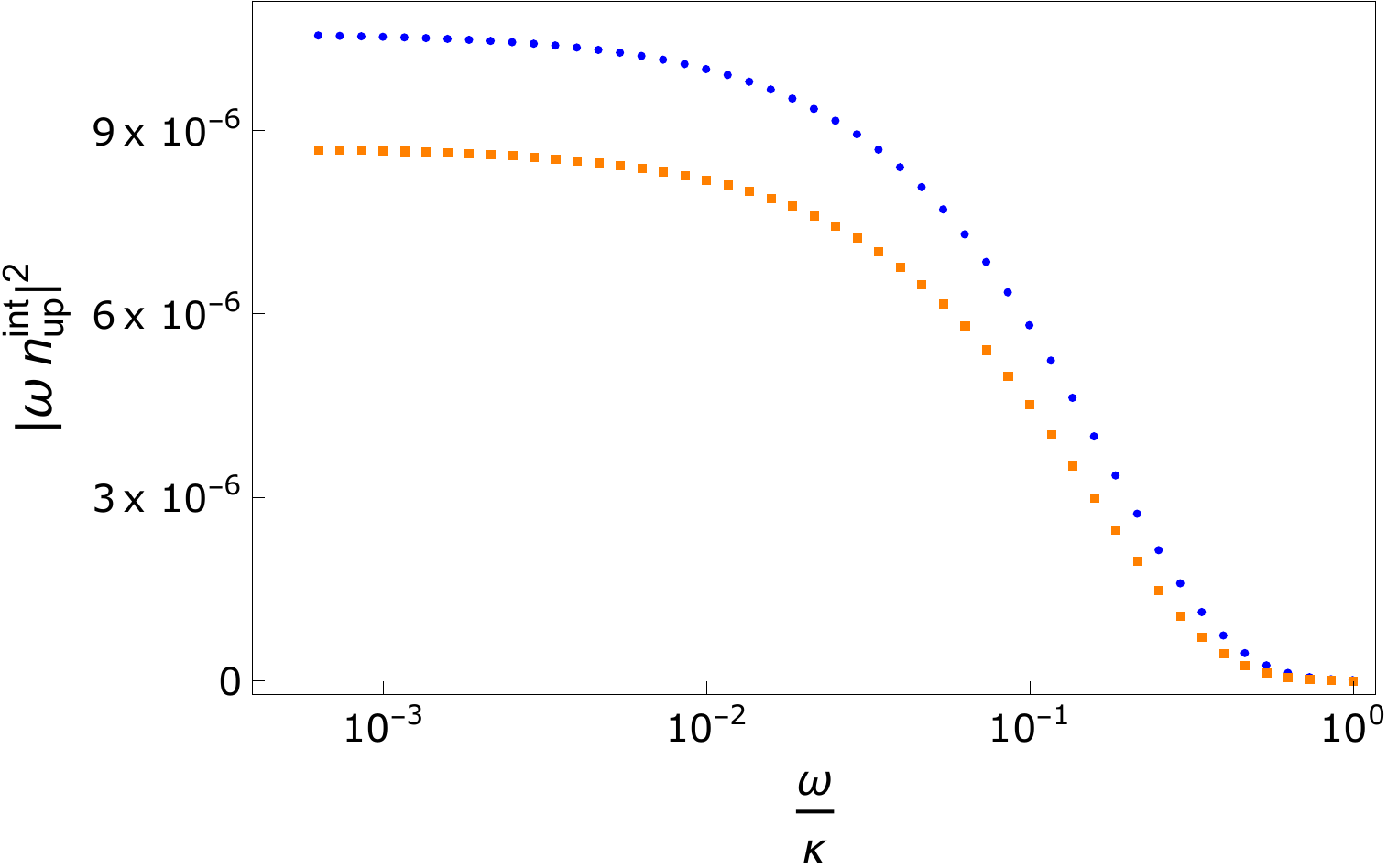}\hspace{.15in}
\includegraphics[width=2.9in]{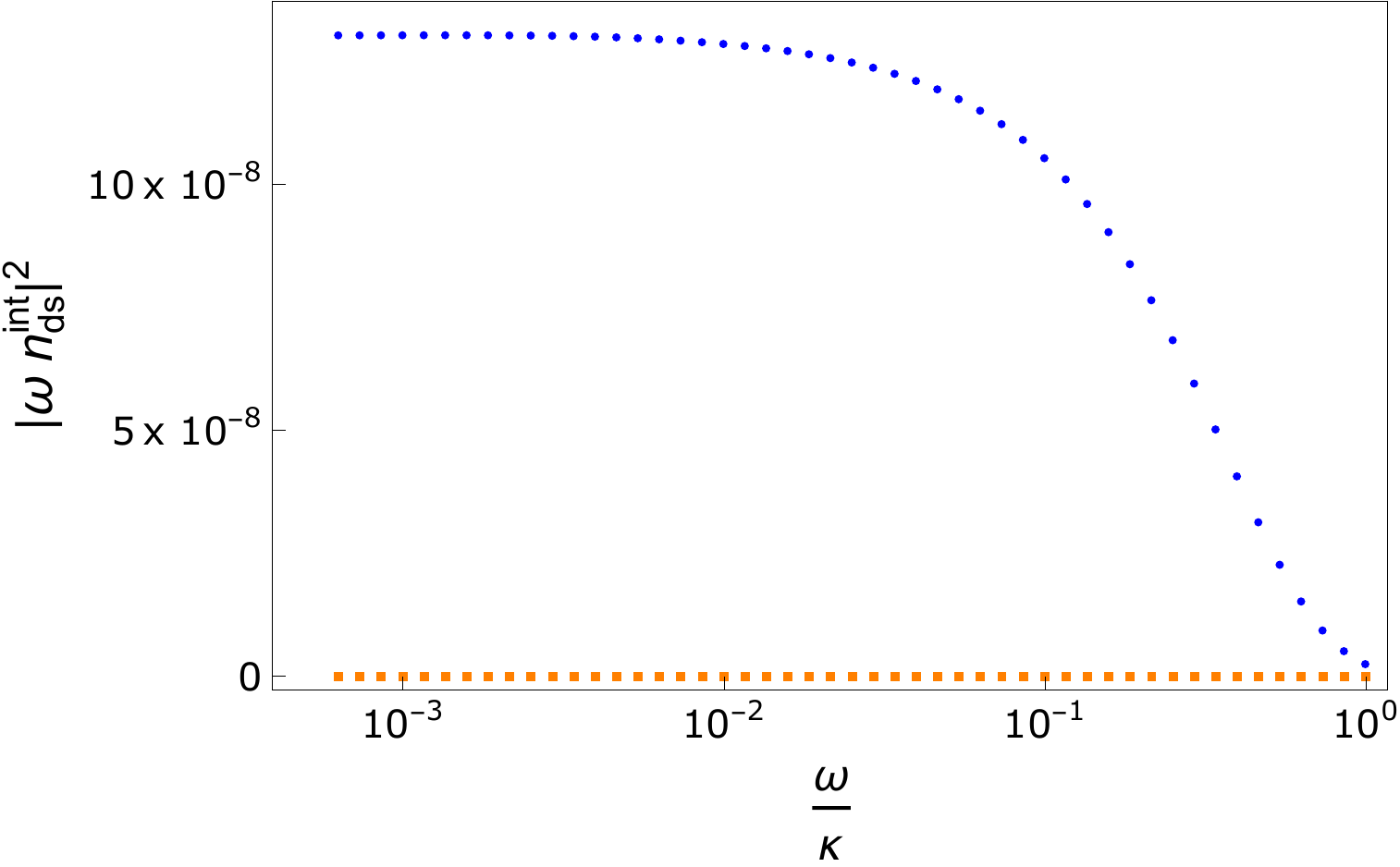}
\caption{\label{Fig:2013InteriorComparisons}Upper: plot of the effective potential for the constant  flow velocity model.  Left: the quantity  $\left| \omega \ n_{\text{up}}^{\text{int}}\right|^2$  vs $\omega$ is shown for the constant flow velocity profile (blue dots)  and for the case where there is no potential (orange squares).  Right:   the quantity  $\left| \omega \ n_{\text{ds}}^{\text{int}}\right|^2$   vs $\omega$ is shown for the  constant flow velocity profile (blue dots)  and for the case where there is no potential (orange squares).   }
\end{figure}

The constant flow velocity model has a speed of sound profile which, for the constants used in the calculations of the HP correlator in Sec.\ref{Sec:ConstantFlowModel},  leads to a nearly antisymmetric effective potential (shown in Fig. \ref{Fig:2013InteriorComparisons}). In this case, the magnitude of the effective potential in the exterior region is only slightly larger than its magnitude in the interior region. The resulting quantities $|\omega n_{\text{up}}^{\text{int}}|^2$ and $|\omega n_{\text{ds}}^{\text{int}}|^2$,  plotted in  Fig. \ref{Fig:2013InteriorComparisons}, do not show a peak and instead are qualitatively similar to $\left| \omega \ n_{\text{up}}^{\text{int}}\right|^2$  in the case where $V_{\text{eff}}=0$. \\

\begin{figure}[h!]	
		\includegraphics[width=2.85in]{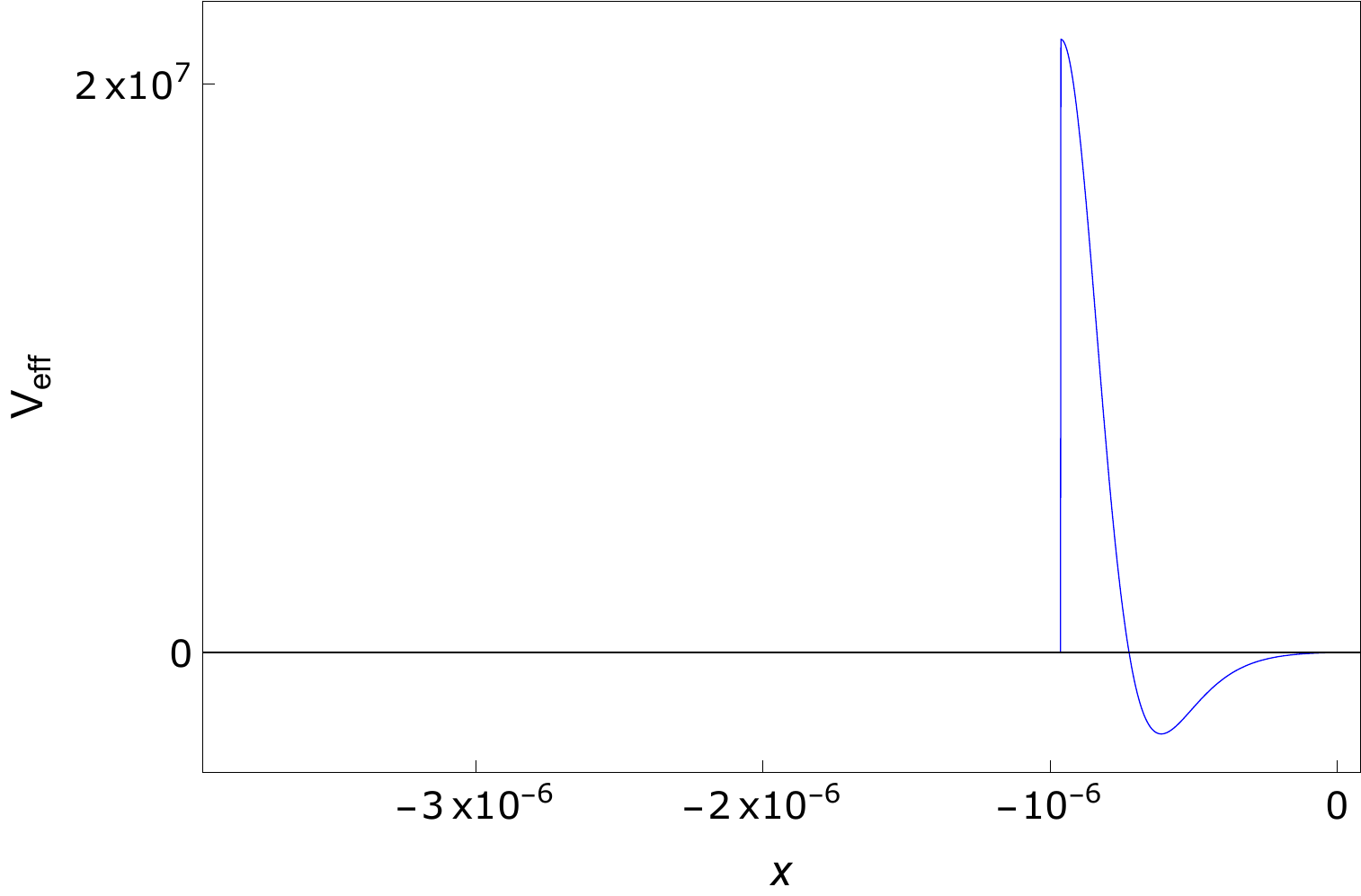}
			\includegraphics[width=2.85in]{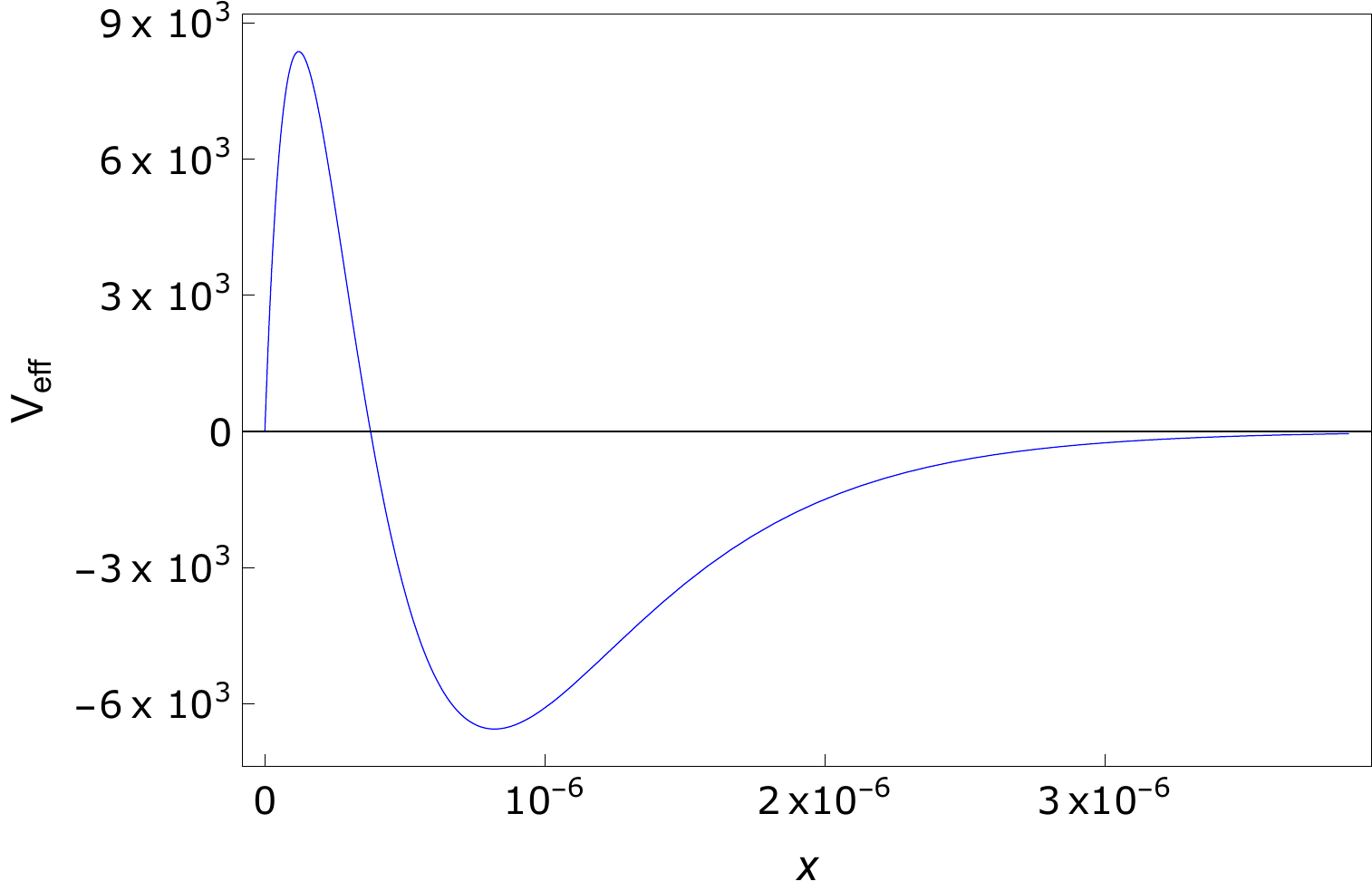}	
			\includegraphics[width=2.85in]{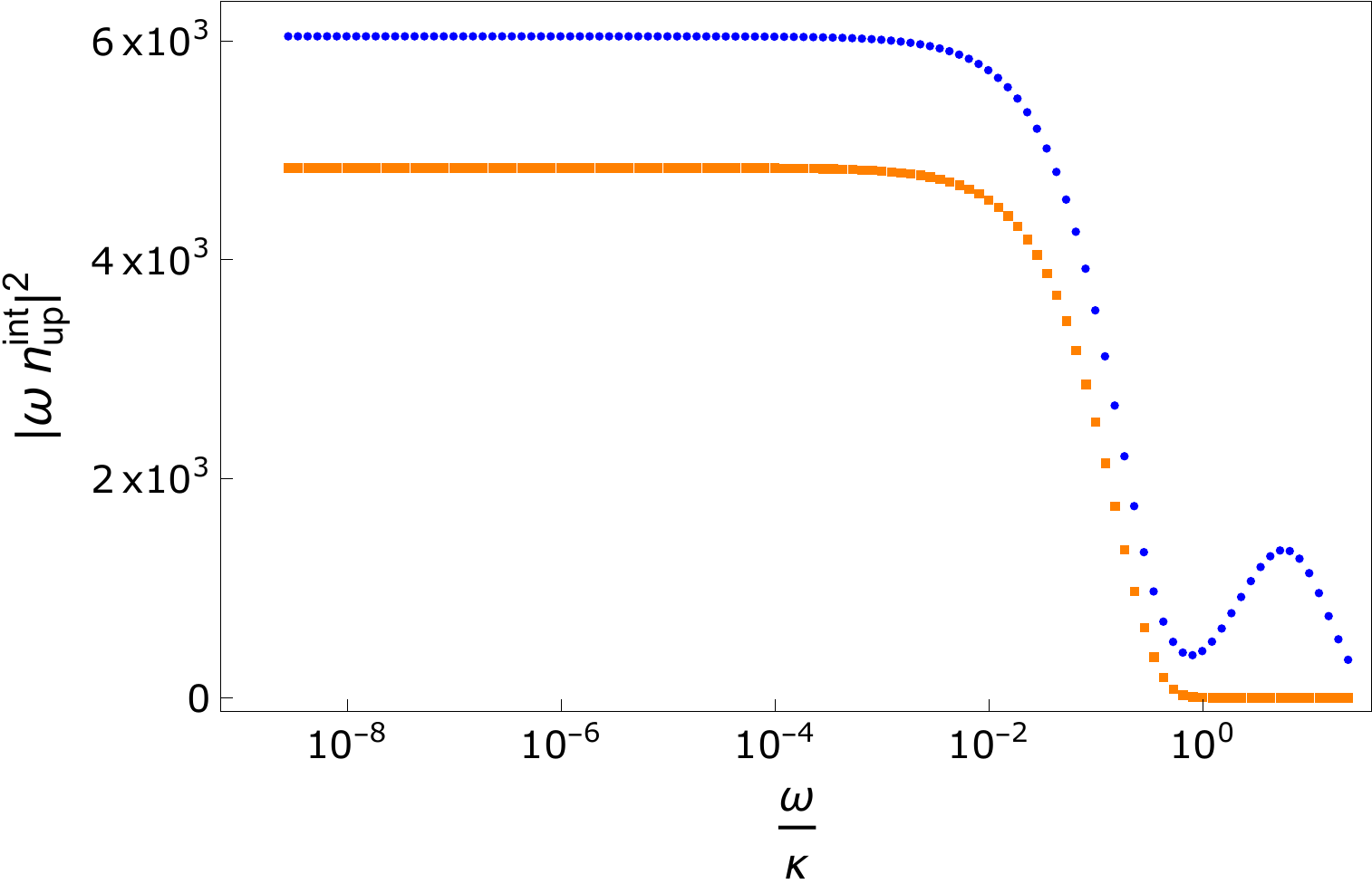}\hspace{.15in}
	\includegraphics[width=2.85in]{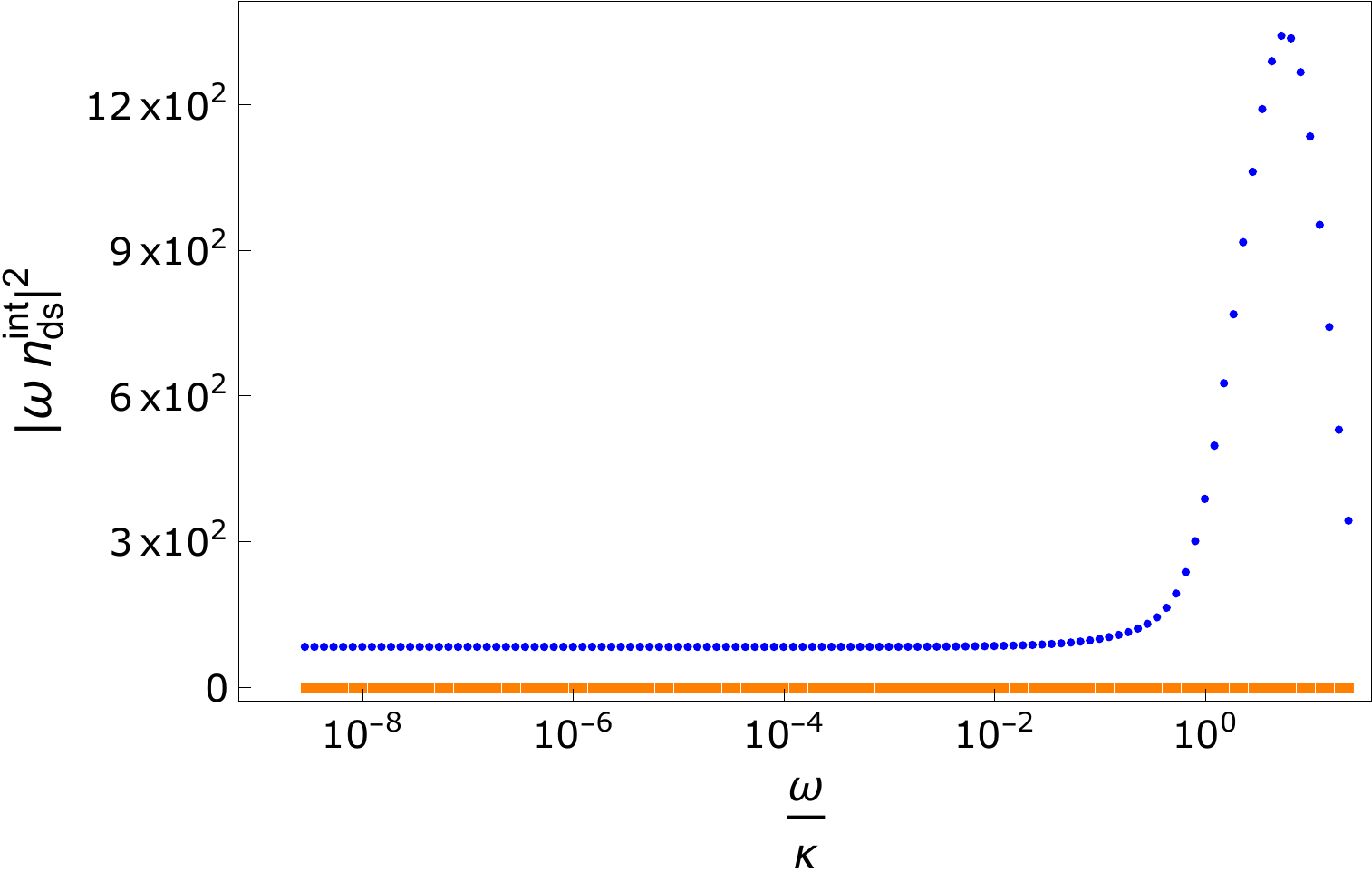}
	\caption{\label{Fig:WaterfallComparisons}Plot of the effective potential for the waterfall model in the interior (upper left) and exterior (upper right) of the analog black hole. The  vertical axes of the upper-right and upper-left plots have different scales
 since the magnitude of the effective potential in the interior is significantly larger than it is in the exterior. Bottom left: the quantity  $|\omega n_{\text{up}}^{\text{int}}|^2$  vs $\omega/\kappa$ is shown for the waterfall model (blue dots)  and for the case where there is no effective potential (orange squares).  Bottom right:  the quantity  $|\omega n_{\text{ds}}^{\text{int}}|^2$  vs $\omega/\kappa$ is shown for the waterfall model (blue dots)  and for the case with no effective potential (orange squares). }
\end{figure}

For the waterfall model, the nature of the profiles for $c(x)$, $v(x)$, and $n(x)$ results in the magnitude of the interior effective potential being much larger than the exterior effective potential, as can be seen in Fig.~\ref{Fig:WaterfallComparisons}. This results in a distinctive peak in the plot of $|\omega n_{\text{up}}^{\text{int}}|^2$, while the plot of $|\omega n_{\text{ds}}^{\text{int}}|^2$ is dominated by the peak as seen in the lower right panel of Fig. \ref{Fig:WaterfallComparisons}.
In the two-delta function potential and waterfall models, one finds for $|\omega n_{\text{up}}^{\text{int}}|^2$ and $|\omega n_{\text{ds}}^{\text{int}}|^2$ what appears to be a peak superimposed on a distribution which is almost thermal.   The structure of  $|\omega n_{\text{up}}^{\text{int}}|^2$ and $|\omega n_{\text{ds}}^{\text{int}}|^2$  for the waterfall model can still be described as a peak superimposed on a thermal distribution, but unlike the two-delta function potential, the peak found in $|\omega n_{\text{ds}}^{\text{int}}|^2$  has a much higher magnitude when compared with the asymptotically constant low frequency region. We also find that in the waterfall model the peaks in both $|\omega n_{\text{up}}^{\text{int}}|^2$ and $|\omega n_{\text{ds}}^{\text{int}}|^2$ appear at higher frequencies in the distribution than was found for the peaks in the two-delta function potential model.
\clearpage

\section{Conclusions\label{Sec:HpCorrelatorConclusions}}
We have studied the
HP correlator, \eqref{Eq:HPCorrelatorGeneric}, and  the interior upstream and downstream phonon numbers,~\eqref{Eq:UpParticleNumberGeneric} and~\eqref{Eq:DsParticleNumberGeneric} for a BEC analog black hole. The mode equation for phonons in the hydrodynamic limit is a wave equation with a potential that depends on the density, flow velocity, and sound speed. In some previous studies, this potential  was neglected for simplicity.   We have shown that the inclusion of this effective potential has a significant impact on the HP correlator and the interior numbers of upstream and downstream phonons in each of the models we have investigated

Three different models were considered. The HP correlator was calculated by solving the mode equation with the effective potential, $V_\text{eff}$, for each model and then comparing the result with the case with no effective potential. The first model has an effective potential consisting of two delta functions, one in the exterior and one in the interior. The behavior of the HP correlator for the two-delta function potential model is quite different from the case where $V_\text{eff}=0$ as the low frequency HP correlator is finite for the two-delta function potential model, whereas it is infrared  divergent if $V_\text{eff}=0$.

A second model  has a constant flow velocity, but a varying sound speed.  In this case the HP correlator is qualitatively similar to the $V_\text{eff}=0$ case. However, at low frequencies, they differ by as much as $8\%$.

The third model, called the waterfall model,  is a solution to the Gross-Pitaevskii  equation for the background density if a step function potential is applied. The resulting profile has a varying sound speed and flow velocity. The HP correlator for this model differs significantly from the case when $V_\text{eff}=0$.  In the low frequency regime, in particular, the HP correlator for the waterfall model is increased by approximately $10\%$ compared to the case when $V_\text{eff}=0$.

We have also calculated the interior UPN and DPN at future null infinity  for the constant flow velocity model and the waterfall model and have also reviewed the results for the two-delta function potential model in \cite{PhysRevD.100.105021}. In the  two-delta function potential model, one finds a peak in both $|\omega n_{\text{up}}^{\text{int}}|^2$ and $|\omega n_{\text{ds}}^{\text{int}}|^2$ when the potential is adjusted so that the interior effective potential  is larger than the exterior. The waterfall model, by its nature,  has an interior effective potential which is much larger in magnitude when compared to the exterior and thus has an easily visible peak in both quantities related to the UPN and DPN. The case with a constant flow velocity does not have a larger effective potential in the interior and no such peak is found in  $|\omega n_{\text{up}}^{\text{int}}|^2$ or $|\omega n_{\text{ds}}^{\text{int}}|^2$.

The same particle production that leads to the peak related to the interior UPN and DPN appears to have a small impact on the HP correlator for the waterfall model. This is only visible when looking at the ratio of the curve with $V_\text{eff}\neq0$ to the curve with $V_\text{eff}=0$ in Fig. \ref{Fig:WaterfallHPCorreleatorComparisons4}. This impact is small enough that we do not expect to see its effect in the current experimental results\cite{Ste2016959,Nature.569.s41586}.

In \cite{PhysRevD.92.024043}, it was shown that there is a relationship between the HP correlator and the Fourier transform of the density-density correlation function when one point is inside and one point is outside the horizon.  A similar relationship was found in~\cite{PhysRevD.92.024043} between the  Fourier transform of the density-density correlation function when both points are inside the horizon and the quantities $|\omega n_{\text{up}}^{\text{int}}|^2$ and $|\omega n_{\text{ds}}^{\text{int}}|^2$.
Given the prominence of the peak in the theoretical
calculation for the waterfall model one could hope to see it in the
experimental data.  Unfortunately, for the experimental configuration in~\cite{Ste2016959}, this does not seem to be
the case~\cite{Jeff-private}.

\acknowledgments
We would like to thank Eric Carlson and Gregory Cook for helpful suggestions regarding the paper and R. A. D. would like thank Jeff Steinhauer for helpful conversations.  R. A. D.   would also like to thank
the University of Valencia, where some of this work was done, for hospitality and he acknowledges partial financial support for the visit from the Paul K. and Elizabeth Cook Richter Memorial Fund.  A.F. acknowledges partial financial support from the
Spanish Ministerio de Ciencia e Innovaci\'on Grant
FIS2017-84440-C2-1-P and from the Generalitat Valenciana Grant
PROMETEO/2020/079.  This work was supported in part by the National
Science Foundation under Grants No.\ PHY-1505875 and No.\ PHY-1912584 to Wake Forest University.

\bibliography{ReferencesGR}

\end{document}